\documentclass[a4paper,11pt]{article}
\usepackage{jheppub} % for details on the use of the package, please
\usepackage{amssymb}
\usepackage{amsmath}
\usepackage{amsthm}
\usepackage{mathrsfs}
\usepackage[numbers]{natbib}
\usepackage{mathtools}
\usepackage{graphicx}
\usepackage{float}
\usepackage{subfigure}
\usepackage{booktabs}
\usepackage{multirow}
\usepackage[colorlinks=true]{hyperref}
\usepackage{physics}

\usepackage[dvipsnames]{xcolor}
\usepackage{ulem}

\title{Kinematic space for quantum extremal surface}
\author[a,b]{An Gong,}
\author[c]{Chong-Bin Chen,}
\author[a,b,d,1]{and Fu-Wen Shu \note{Corresponding author.}}
\affiliation[a]{Department of Physics, Nanchang University, Nanchang, 330031, China}
\affiliation[b]{Center for Relativistic Astrophysics and High Energy Physics, Nanchang University, Nanchang, 330031, China}
\affiliation[c]{Department of Physics, Kobe University, Kobe 657-8501, Japan}
\affiliation[d]{Center for Gravitation and Cosmology, Yangzhou University, Yangzhou, China}
\emailAdd{shufuwen@ncu.edu.cn}

\abstract{This paper investigates the entanglement entropy inequality and explores the presentation of mutual information and conditional mutual information in kinematic space. Specifically, we examine the regions within kinematic space responsible for computing these physical quantities, enabling a more intuitive understanding of the entanglement entropy inequality. Building upon this, we employ the concept of double holography to analyze the properties of the entanglement inequality in any given region. By utilizing kinematic space, we calculate the contribution of the bulk to the holographic entanglement entropy in double holography. In conclusion, we establish that kinematic space substantiates a conjecture, namely that the entanglement entropy of an entire region can be expressed as a linear combination of the entanglement entropy of a single interval within the entangled region.}

\begin{document}
\maketitle
\flushbottom

\section{Introduction}

The holographic viewpoint has garnered attention for decades since the development of string theory in the last century. Holography posits that (d+2)-dimensional quantum gravity theory exhibits significantly fewer degrees of freedom than previously hypothesized. In fact, its degrees of freedom are equivalent to those of a (d+1)-dimensional quantum many-body theory \cite{1,2,Susskind:1994vu}. This discovery originates from the observation that the entropy of a black hole is not proportional to its volume but to the area $\Sigma$ of its event horizon, known as the Bekenstein-Hawking formula \cite{3}. An exemplary illustration of the holographic principle is the celebrated AdS (anti-de Sitter)/CFT (conformal field theory) correspondence \cite{4,5,6,7}. This duality signifies a significant advancement in our understanding of string theory and quantum gravity.

However, despite years of research on AdS/CFT, the nature of the AdS/CFT duality remains elusive. Specifically, we are yet to determine which region of AdS is responsible for specific information in the dual CFT. To address this longstanding issue, it was realized that the holographic principle can be comprehended through the concept of holographic entanglement entropy. This principle states that the entanglement entropy of a subregion in the boundary CFT is proportional to a codimension-two minimum surface known as the Ryu-Takanayagi surface (RT surface) in the AdS bulk \cite{23,24}. Mathematically, this relationship is expressed as:

\begin{equation}\label{rt}
S_{RT}(A)=\frac{A^{(d)} (\gamma_A)}{4G^{(d)}},
\end{equation}
where, $A^{(d)} (\gamma_A)$ denotes the area of the minimal surface $\gamma_A$ that is homologous to region $A$, while $G^{(d)}$ represents the Newton constant in $d$ dimensions.

It was later discovered that the holographic entanglement entropy formula \eqref{rt} should take into account the entanglement entropy of the matter field in the bulk, leading to a holographic entanglement entropy formula with a first-order quantum correction term \cite{27,28}. This formula was further generalized to incorporate higher-order quantum corrections by replacing the RT surface with the quantum extremum surface (QES), as demonstrated in \cite{29}. The introduction of QES revitalized investigations into the black hole information paradox \cite{Hawking:1976ra}, which has perplexed the physics community for several decades. Inspired by the formulation of QES, the concept of the entanglement island prescription was proposed \cite{Penington:2019npb,Almheiri:2019psf,Almheiri:2019hni}. Remarkably, it was discovered that the Page curve \cite{Page:1993wv,Page:2013dx}, which describes the unitary evolution of black hole evaporation, can be replicated in various models \cite{Almheiri:2019yqk,Chen:2019uhq,Almheiri:2019psy,Gautason:2020tmk,Hashimoto:2020cas,Anegawa:2020ezn,
Hartman:2020swn,Hollowood:2020cou,Alishahiha:2020qza,Bak:2020enw,Dong:2020uxp,Krishnan:2020fer,Chen:2020jvn,Ling:2020laa,Matsuo:2020ypv,Goto:2020wnk,Caceres:2020jcn,Karananas:2020fwx,Wang:2021woy,Kim:2021gzd,Lu:2021gmv,Yu:2021cgi,Arefeva:2021kfx,He:2021mst,Arefeva:2022guf,Arefeva:2022cam,Geng:2020qvw,Geng:2020fxl,Geng:2021hlu,Saha:2021ohr,Ahn:2021chg,Krishnan:2020oun,Omidi:2021opl,Gan:2022jay,Matsuo2021,Ageev:2022hqc,Guo:2023gfa,Chen:2020uac,Chen:2020hmv}.

Another significant achievement in exploring the potential connections between information theory and the bulk spacetime in the AdS/CFT correspondence is the concept of kinematic space \cite{Czech:2015qta}. Kinematic space is an auxiliary Lorentzian geometry formed by mapping geodesics that are anchored on the boundary of the AdS bulk. In essence, kinematic space can be understood as a space composed of geodesic lines, and the coordinates required to determine a geodesic line correspond to a point in kinematic space. By integrating over a specific region in kinematic space, one can obtain geometric quantities in the original space, such as the length of a geodesic line. This concept offers a fresh perspective on the interplay between information theory and the geometric description of a CFT state. Specifically, it encodes geometric variables, such as the lengths of bulk curves, into boundary quantities, which can be interpreted entropically in the CFT. Furthermore, kinematic space proves advantageous in describing the causal structures of intervals. It has been demonstrated in \cite{Czech:2015kbp} that the multiple entanglement renormalization ansatz (MERA) tensor network is more suitably described in terms of kinematic space rather than the spatial slice of the holographic AdS geometry. In recent years, the applications of kinematic space have been explored in various aspects. For example, the OPE blocks in the CFT can be regarded as dynamical scalar fields propagating in the kinematic space \cite{Czech:2016xec}. The OPE block is an important quantity for studying the bulk physics, such as the local operator reconstruction. With this picture, it's revealed that the dynamics of the Entanglement entropy in the kinematic space is closely related to the Einstein's equation in the bulk \cite{Czech:2016tqr,deBoer:2015kda,Asplund:2016koz,Mosk:2016elb}. Also, the application of kinematic space in the quantum bit threads of holographic entanglement entropy \cite{Chen:2018ywy}, and holographic complexity \cite{Czech:2017ryf,Chen:2018ody} are also demonstrated.

In this work, our aim is to establish connections between the concept of QES and kinematic space, utilizing the framework of double holography. Double holography suggests that a holographic gravity dual to a boundary conformal field theory (BCFT) involves two layers of holography, leading to three fundamentally equivalent descriptions \cite{Almheiri:2019hni,Almheiri:2019psy,Chen:2020uac,Chen:2020hmv}. Given the significant advancements in the understanding of QES in the context of black hole information theory, and the profound links between kinematic space and information theory, exploring potential connections between QES and kinematic space has the potential to provide novel insights into both the black hole information paradox and quantum information theory. Indeed, our findings suggest a correlation between the configuration of islands and the CFT. In addition, as an application, we discover that we can utilize kinematic space to investigate holographic entanglement inequalities. By employing double holography, we can even derive holographic entanglement entropy inequalities for disjoint intervals.

The structure of this paper is as follows. In the subsequent section, we provide a brief overview of integral geometry and the concept of double holography, which extends the notion of kinematic space to encompass boundary intervals. In Section 3, we propose an extension of the concept of kinematic space to incorporate QES through the framework of double holography. Then in section 4 we utilize kinematic space to examine holographic entanglement inequalities, offering an intuitive understanding from the perspective of kinematic space. In this section we also applies the double holographic viewpoint to derive holographic entanglement entropy inequalities for disjoint intervals. In Section 5, we present the proof of a conjecture concerning the contour function using kinematic space. The final section concludes with a discussion and outlook for future research.

\section{Kinematic space and double holography: an overview}
In this section, we would like to give a brief review on the basic concept of kinematic space and the double holography. 
%In the concept of integral geometry, we can obtain the kinematic space of different geometric surfaces. In Section 2.2, we introduced three different descriptions of holographic theory, which are the dual holographic views.In Chapter 3 we use kinematic space to discuss holographic entanglement inequalities, which can be seen intuitively from kinematic space.

\subsection{Kinematic space}
Kinematic space serves as an auxiliary dual geometry, where its points correspond to geodesic lines in the original AdS space. Its primary function is to mediate the translation between the language of information theory in the boundary and the language of geometry in the bulk for a given manifold \cite{Czech:2015qta}. Therefore, kinematic space can be seen as a translator that converts the boundary language of information theory into the bulk language of geometry. One illustrative example is the encoding of geometric variables into boundary quantities, which can be interpreted in terms of entropy in the context of CFT. To illustrate this point, let us consider the length of a bulk curve. For a closed and smooth convex curve $\gamma$ in the Euclidean plane, the length of the curve is given by
\begin{equation}\label{c1}
	\textrm{length of $\gamma$} = \frac{1}{4} \int_K \omega(\theta, p)\, n_\gamma(\theta,p) \, ,
\end{equation}
where $\theta\in[0,2\pi]$ represents the polar angle on the plane, $p\in[-\infty,\infty]$ denotes the distance of the straight line from the origin, and $n_\gamma(\theta,p)$ represents the intersection number with $\gamma$. The integration on the right-hand side of the equation is performed over kinematic space $K$, which is the space comprising all directional geodesics in the Euclidean plane. The measure of integration, known as the Crofton form, is given by
\begin{equation}
	\omega = dp \wedge d\theta \, .
\end{equation}
As can be observed from the Crofton formula, it allows the transformation of the calculation of the curve's length in the original space to the calculation of the volume in the geodesic space.

The Crofton formula \eqref{c1} can be extended to holographic spactimes, which are of particular interest in the present paper. It turns out for a static time slice of a holographic spacetime, for instance, the hyperbolic plane for pure AdS$_3$ case, we have
\begin{equation}\label{length}
	\textrm{length of $\gamma$} = \frac{1}{4} \int_K \omega(u,v)\, n_\gamma(u,v) \, .
\end{equation}
where $u$ and $v$ are the endpoints of the geodesic line anchored to the cutoff boundary of the hyperbolic plane, and the integral measure now becomes \cite{Czech:2015qta}
\begin{eqnarray}\label{c2}
	\omega(u,v) &=&\frac{\partial^2 L(u,v)}{\partial u\partial v}du \wedge dv\\
	&=& \frac{1}{2 \sin^2\left(\frac{v-u}{2}\right)}\, du \wedge dv\,,
\end{eqnarray}
where $L(u,v)$ is the length of a geodesic connecting the boundary points $u$ and $v$. Combining  \eqref{rt} and \eqref{c2}, it is straightforward to derive
\begin{eqnarray}\label{c3}
	\omega(u,v) =\frac{\partial^2 S(u,v)}{\partial u\partial v}du \wedge dv,
\end{eqnarray}
if we set $4G^{(d)}=1$. In other words, the integral measure encodes the geometry of the Euclidean and hyperbolic planes, which have deep physical meaning in the AdS$_3$/CFT$_2$ duality, where the Ryu-Takayanagi proposal identifies bulk geodesic lengths with boundary entanglement entropies.
%We hope to obtain the integral measure of the kinematic space through the FLM formula or QES\textcolor{red}{(You need to write down the full name of FLM and QES when they first appear in you article)}, so as to combine the concept of integral geometry Promote the application.

\subsection{Double Holography}
The notion of double holography has gained renewed interest in recent years, particularly in its applications to the context of the black hole information paradox \cite{Almheiri:2019hni}. It was first discovered in the Karch-Randall model \cite{Karch:2000gx, Karch:2000ct}, which is based on the AdS/BCFT correspondence. This correspondence establishes a holographic duality between an asymptotically AdS bulk, where part of the asymptotic AdS boundary is replaced by an AdS brane, and a BCFT whose boundary is located at the intersections of the bulk brane and the asymptotic boundary.

\begin{figure} \centering 
\subfigure[The BCFT description] { \label{fig:a} 
\includegraphics[width=0.25\linewidth]{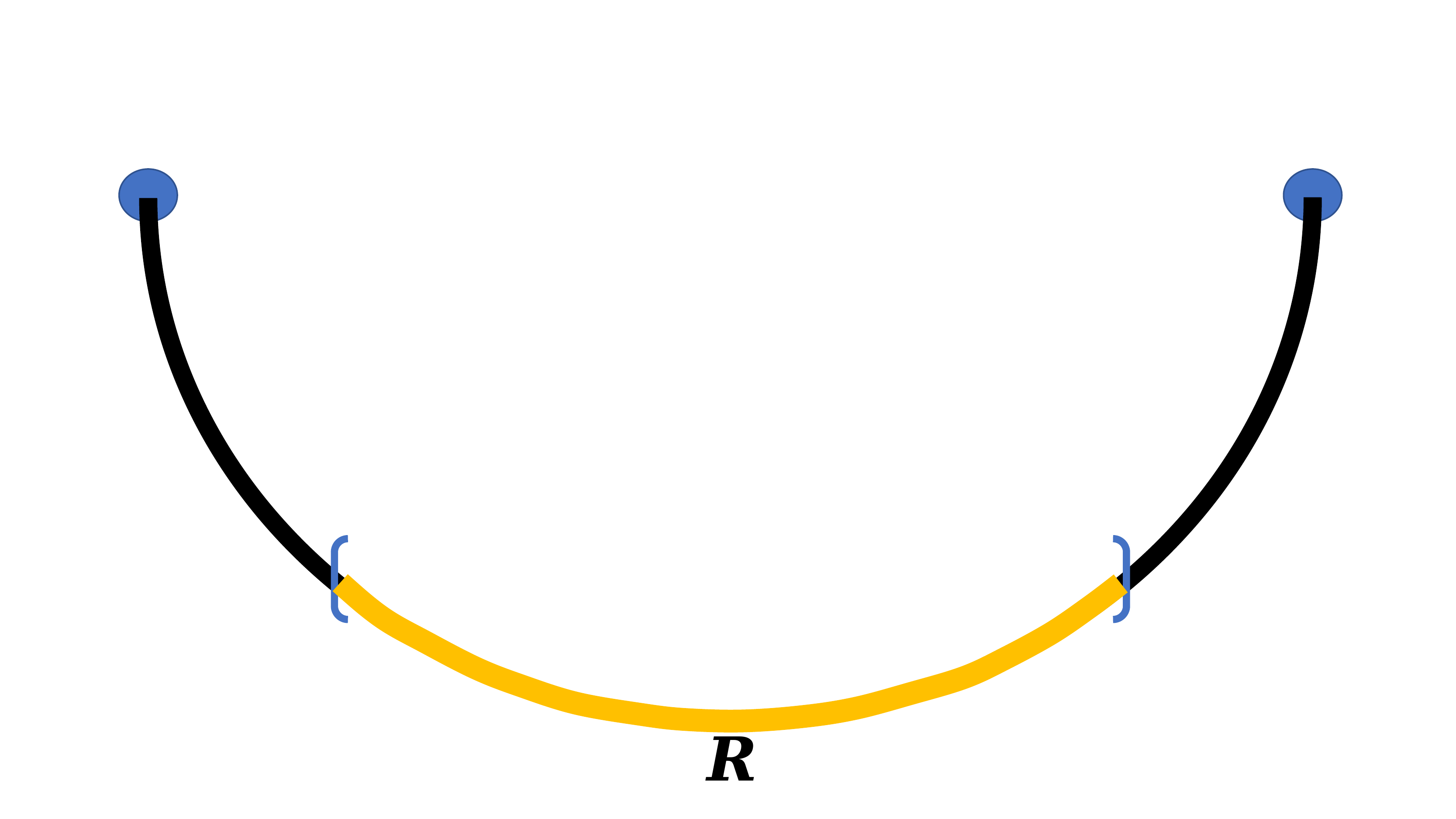} 
} 
\subfigure[The AdS description] { \label{fig:b} 
\includegraphics[width=0.34\linewidth]{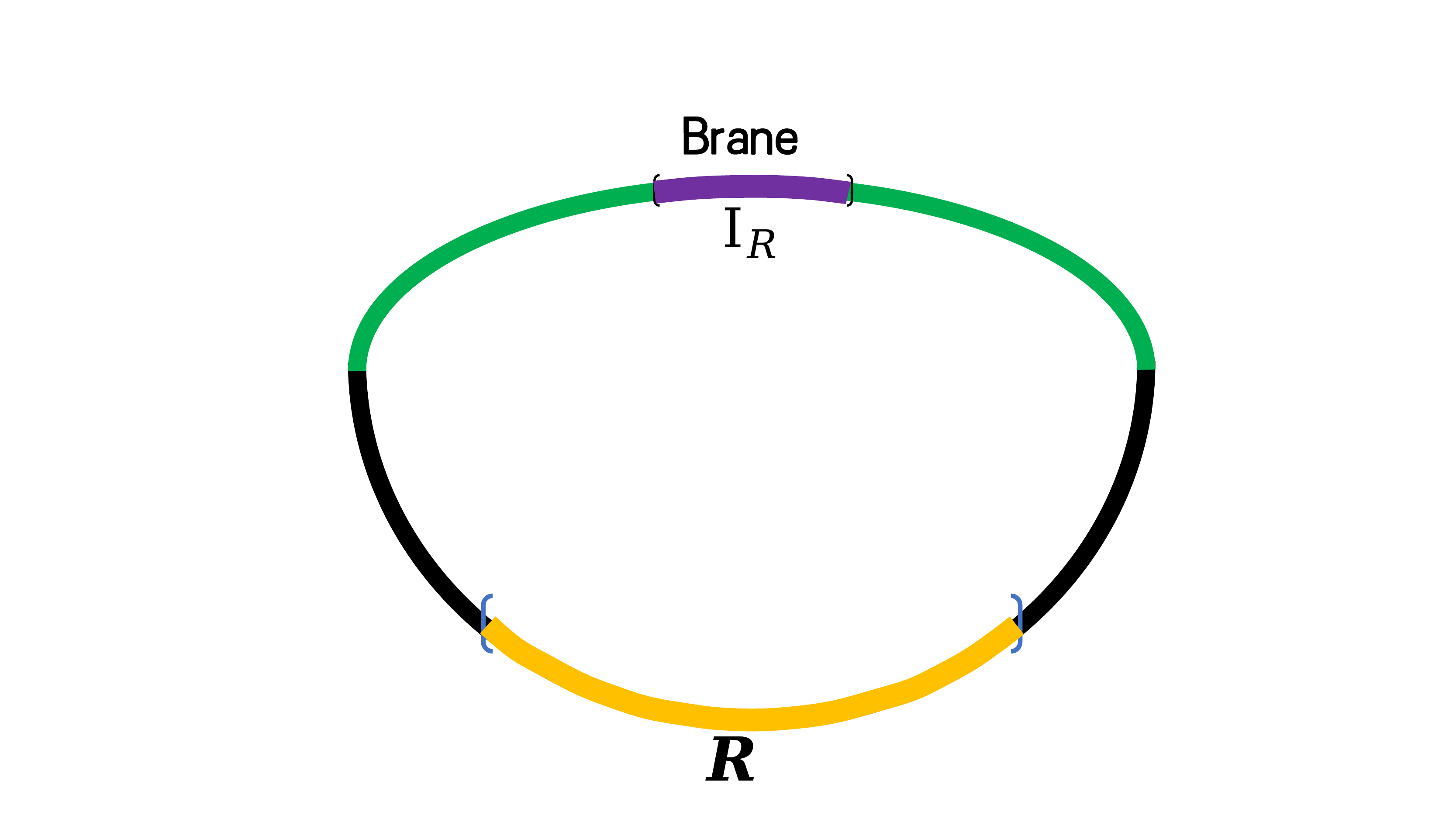} 
} 
\subfigure[The EOW description] { \label{fig:c} 
\includegraphics[width=0.34\linewidth]{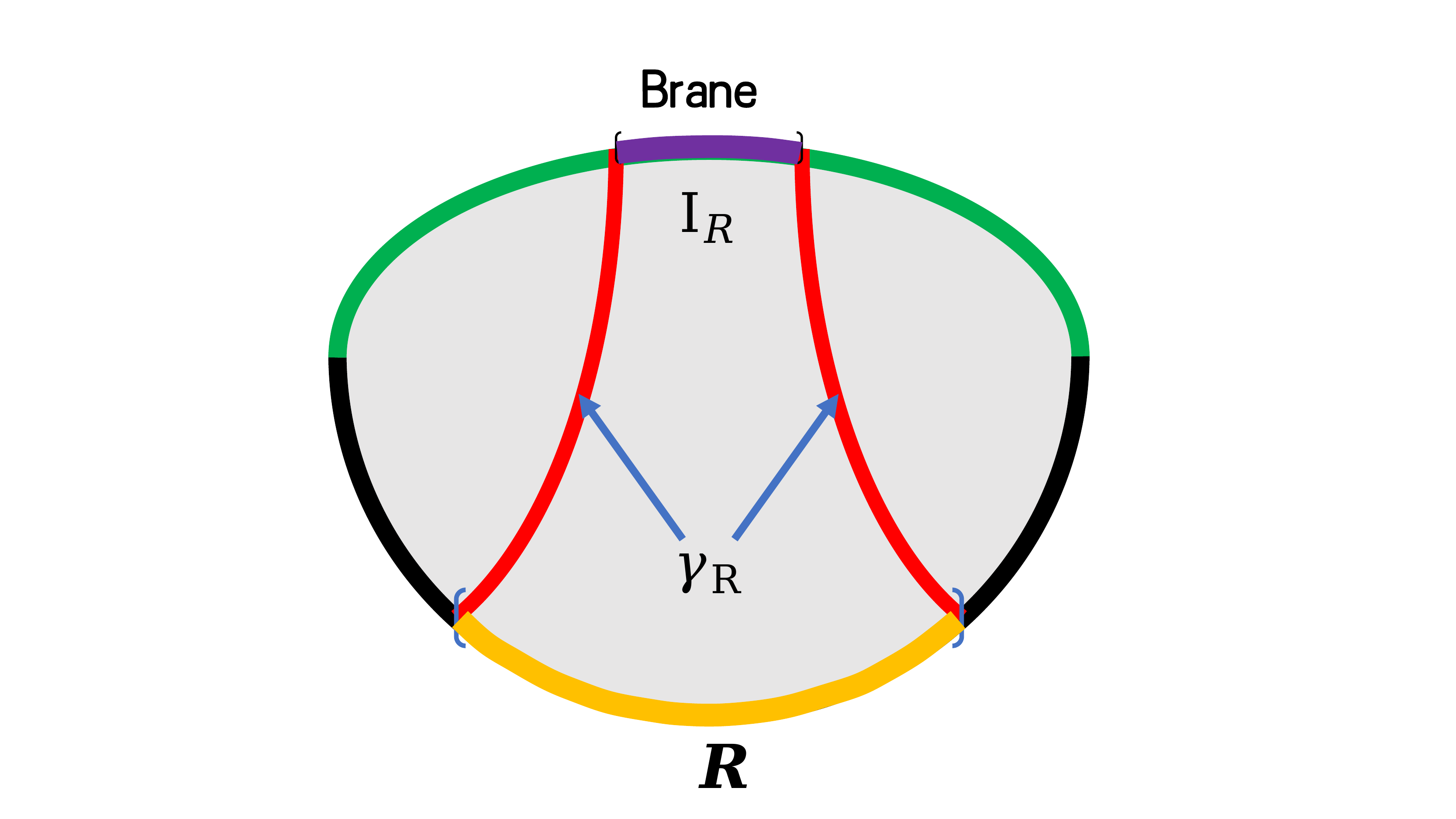} 
} 
\caption{ Three equivalent descriptions of the AdS/BCFT. (a) The BCFT description.   A $d$-dimensional BCFT (the black solid line) where the CFT coupled to $(d-1)$-dimensional boundary defects (the blue points). $R$ (the orange line) is a subsystem of the BCFT. (b) The AdS description. Description in terms of an asymptotically $d$-dimensional AdS geometry where parts of the asymptotic boundary are regulated by a brane (the green line). The purple line refers to the entanglement island of the corresponding region $R$. (c) The EOW description. Description in terms of a $(d+1)$-dimensional AdS spacetime (the shaded region) with an EOW brane.} 
\label{fig1} 
\end{figure}

According to double holography, an AdS geometry dual to a boundary CFT (i.e., the AdS/BCFT correspondence) actually has not just two but three equivalent descriptions. Specifically, the AdS/BCFT correspondence has two dual descriptions: the boundary description, where the CFT is coupled to codimension-one boundary defects (referred to as the BCFT description), and the bulk description, where the CFT is coupled to a brane replacing the boundary defects in the first description (referred to as the AdS description). However, it has been found that when the AdS brane is very close to the asymptotic boundary, the holographic system admits a third description beyond the usual AdS/BCFT duality. In this case, there is an equivalent description in terms of gravity on a higher-dimensional Randall-Sundrum brane, which serves as the ``end of the world'' (EOW) brane, as shown in Fig. \ref{fig1}. Throughout this paper, we refer to this description as the EOW description.

Now let us consider a subregion $R$ in the BCFT, the entanglement entropy $S(R)$ can be calculated in three different ways in terms of these three descriptions.
In terms of the BCFT description, the entropy of $R$ is calculated by boundary conformal field theory. On the other hand, in terms of the AdS description, the entropy of $R$ can be calculated by using the QES formula including contributions from the island $I_R$ of $R$
   \begin{equation}\label{flm}
  	S_{\text{QES}}(R)=\text{min}\left \{\text{ext}\left [ \frac{A^{(d)}(\partial I_{R}  ) }{4G^{(d)} }+S_{\text{bulk}}(R\cup I_{R} ) \right ]  \right \},
   \end{equation}
where ``ext'' means that we take an extremal value over all possible $I_R$ and ``min'' represents taking the minimal one of the extremums. While in terms of the EOW description, one can compute the entropy by using the RT surface ($\gamma_R$ in Fig. \ref{fig1}) which is anchored to subregion $R$ and its island $I_R$.
  \begin{equation}
  S_{\text{QES}}(R)={\text{min}}\left \{ \text{ext}\left [ \frac{A^{(d)} (\partial I_{R} )}{4G^{(d)} }+\frac{A^{(d+1)} (\gamma _{R})}{4G^{(d+1)} }  \right ]  \right \},\label{sqes0}
  \end{equation}
where ``ext'' means that we take an extremal value over all possible $\gamma_R$.

\section{The kinematic space of quantum extremum surface}
\begin{figure}
	\centering
	\includegraphics[width=1\textwidth]{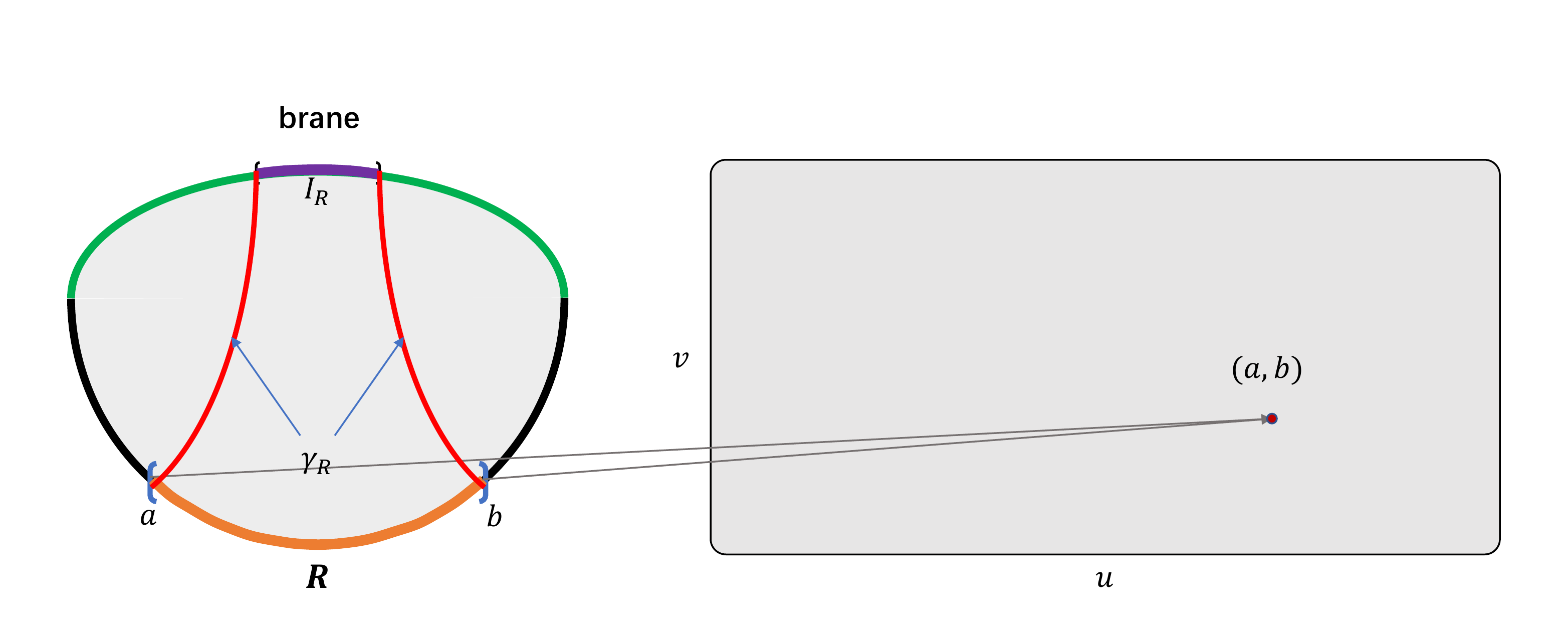}
	\caption{Sketch of kinematic space for QES: The boundary interval R with endpoints ($a,b$) is mapped to a point in the kinematic space.}
	\label{dhg map to k}
\end{figure}

In this section, we propose an extension of the concept of kinematic space to incorporate QES through the framework of double holography. The key distinction is that, instead of mapping RT surfaces, we now consider the mapping of QESs onto the kinematic space. This novel approach enhances our understanding of the entanglement structure and holography within the AdS/CFT correspondence. As depicted in Fig. \ref{dhg map to k}, let us consider a boundary interval R with endpoints (a,b). Within the holographic dual, there exists a unique homologous QES surface, denoted as $\gamma_R$, which corresponds to a specific point in the kinematic space. %This can be viewed as k space for two disjoint subregions in $AdS_{d+1}$ bulk. ..................

The integral measure of the kinematic space composed of geodesics is
\begin{equation}\label{measure}
	\omega(u,v) = \frac{\partial^2 S_{\text{QES}}(u,v)}{\partial u\, \partial v} \, du \wedge dv\,,
\end{equation}
where $S_{\text{QES}}(u,v)$ represents the length of the geodesic with its endpoints located on the cutoff surface, and $u$ and $v$ denote the positions of the endpoints on the boundary. When the QES formula \eqref{sqes0} is applicable, $S_{\text{QES}}(u,v)$ corresponds to the entanglement entropy of the interval $(u,v)$.

We should note that the integral of $\omega(u,v)$ no longer represents the length of the geodesics, as it did in \eqref{length}. This is because, according to \eqref{flm}, the integral now includes not only the area term but also the bulk entropy.

To demonstrate that $\omega(u,v)$, as defined in \eqref{measure}, serves as the integral measure of kinematic space, it is important to establish its positive definiteness. It turns out this can be achieved by considering the requirement of strong subadditivity of entanglement entropy $S_{\text{QES}}(u,v)$. The strong subadditivity of entanglement entropy is formulated as follows:
\begin{equation}\label{strongsubadd}
	S_{\text{QES}}(AB) + S_{\text{QES}}(BC) - S_{\text{QES}}(B) - S_{\text{QES}}(ABC) \geq 0\,,
\end{equation}
where $A$, $B$ and $C$ are three disjoint boundary intervals. Assuming the three boundary intervals are:
\begin{equation}
	A = (u - du, u) \qquad {\rm and} \qquad B = (u,v) \qquad {\rm and} \qquad C = (v, v+dv).\
\end{equation}
From the strong subadditivity, we know:
\begin{equation}
	\begin{split}
	&S_{\text{QES}}(u-du, v) + S_{\text{QES}}(u,v+dv) - S_{\text{QES}}(u,v) - S_{\text{QES}}(u-du, v+dv) 
	\\
	&\approx \frac{\partial^2 S_{\text{QES}}(u,v)}{\partial u \,\partial v} \,du\, dv \geq 0.
	\end{split}
\end{equation}
In later sections, we will explore how to utilize the non-negativity of the integral measure to prove the QES entropy inequalities. This can be graphically represented in the kinematic space defined by the entanglement entropy.

Similarly, we can also define the conditional entropy and mutual information when the QES formula is applicable. The expressions for conditional entropy and mutual information are as follows:
\begin{equation}
S_{\text{QES}} (A|B) = S_{\text{QES}}(AB)-S_{\text{QES}}(B),
\end{equation}
\begin{equation}
I_{\text{QES}}(A, B) = S_{\text{QES}}(A)-S_{\text{QES}}(A|B).
\end{equation}
From the definitions of conditional entropy and mutual information, we can derive the definition of conditional mutual information:
\begin{equation}
	I_{\text{QES}}(A,C | B) \equiv S_{\text{QES}}(A|B) - S_{\text{QES}}(A|BC) = I_{\text{QES}}(A, BC) - I_{\text{QES}}(A, B). 
\end{equation}
Conditional mutual information represents the volume integral of an infinitesimal rectangle in kinematic space. By leveraging the additivity of volumes, we can integrate over the corresponding regions in kinematic space to calculate the conditional mutual information. The relationship between conditional mutual information for different boundary intervals is reflected in kinematic space as an integral over a specific rectangular area, allowing for a straightforward interpretation. The chain rule of conditional mutual information can be intuitively derived from the kinematic space, providing insights into more complex scenarios.
\begin{equation}
	I_{\text{QES}}(A, CD | B) = I _{\text{QES}}(A, C | B) + I_{\text{QES}}(A, D | BC).
\end{equation}

%\subsection{Geometry in kinematic space}
%In the second chapter, we reviewed the basic concepts of integral geometry, and learned how to obtain the length of the curve through the Crofton formula. In the third chapter, we generalized the concept of integral geometry to obtain the kinematic space when the QES formula is applicable. Explain the physical meaning of specific points, lines, and surfaces in kinematic spaces.

\section{Application to proof of QES entropy inequalities}
In this section, we will utilize the concepts introduced in the previous section to derive holographic entanglement entropy inequalities as an application. By leveraging the notion of kinematic spaces, we will demonstrate how these inequalities can be obtained in an intuitive manner. Our discussion is divided into two parts: the AdS description and the EOW description.

\subsection{Proof of QES entropy inequalities in the AdS description}
In this subsection, we aim to demonstrate how QES entropy inequalities can be proven in the AdS description using the concept of kinematic space. To begin, we need to review the definition of geometric quantities, such as points and point curve in kinematic spaces. A point in kinematic space: the boundary interval $(u, v)$ is mapped to a point in kinematic space. Point curve in kinematic space: point $P$ on the boundary is the common endpoint of the boundary interval $(u, v)$, and the point curve of point $P$ in kinematic space is the curve drawn in kinematic space by all boundary intervals whose common endpoint is point $P$.

Using the concept of point curves, we can calculate the entanglement entropy between two points on the boundary. Let $p_a$ represent the point curve of point $a$ and $p_b$ represent the point curve of point $b$. The area enclosed by the point curves of these two points in kinematic space is denoted as $p_a\Delta p_b$. By integrating over $p_a\Delta p_b$, we can determine the entanglement entropy of the boundary interval $(\theta_a, \theta_b)$.
\begin{equation}\label{sqes}
	S_{\text{QES}}(\theta_a, \theta_b)= \frac{1}{4} \int_{{p}_{a}\Delta  {p}_{b}}\omega.
\end{equation}
\begin{figure}[t!]
	\centering
	\includegraphics[width=0.5\textwidth]{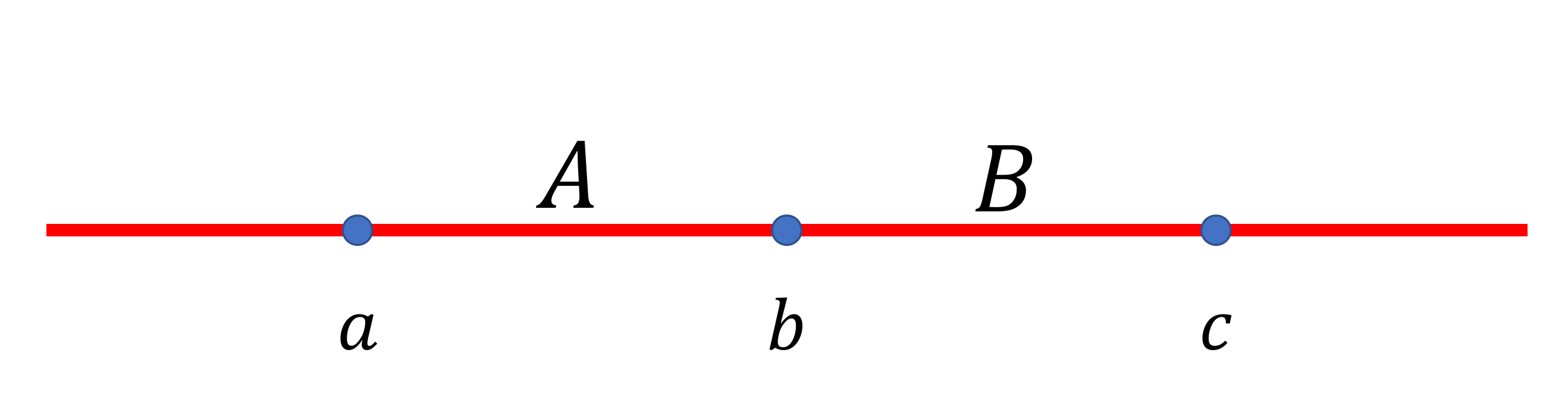}
	\caption{Subsystems on the boundary CFT, where $A$ and $B$ are two joint boundary intervals.}\label{2intervals}
\end{figure}

\subsubsection{Subadditivity} 

Subadditivity claims that for two non-overlapping boundary regions $A$ and $B$, the following inequality for the related entanglement entropies holds:
\begin{equation} 
S_{\text{QES}}(A)+S_{\text{QES}}(B)\ge S_{\text{QES}}(AB).
\end{equation} 
where $A$ and $B$ are two boundary joint intervals, $a$ and $b$ are the endpoints of the boundary interval A, $b$ and $c$ are the endpoints of the boundary interval $B$, and the point curves of points $a$, $b$ and $c$ are shown in the Fig. \ref{2intervals}.

\begin{figure} \centering 
\subfigure[] { \label{fig:a} 
\includegraphics[width=0.35\linewidth]{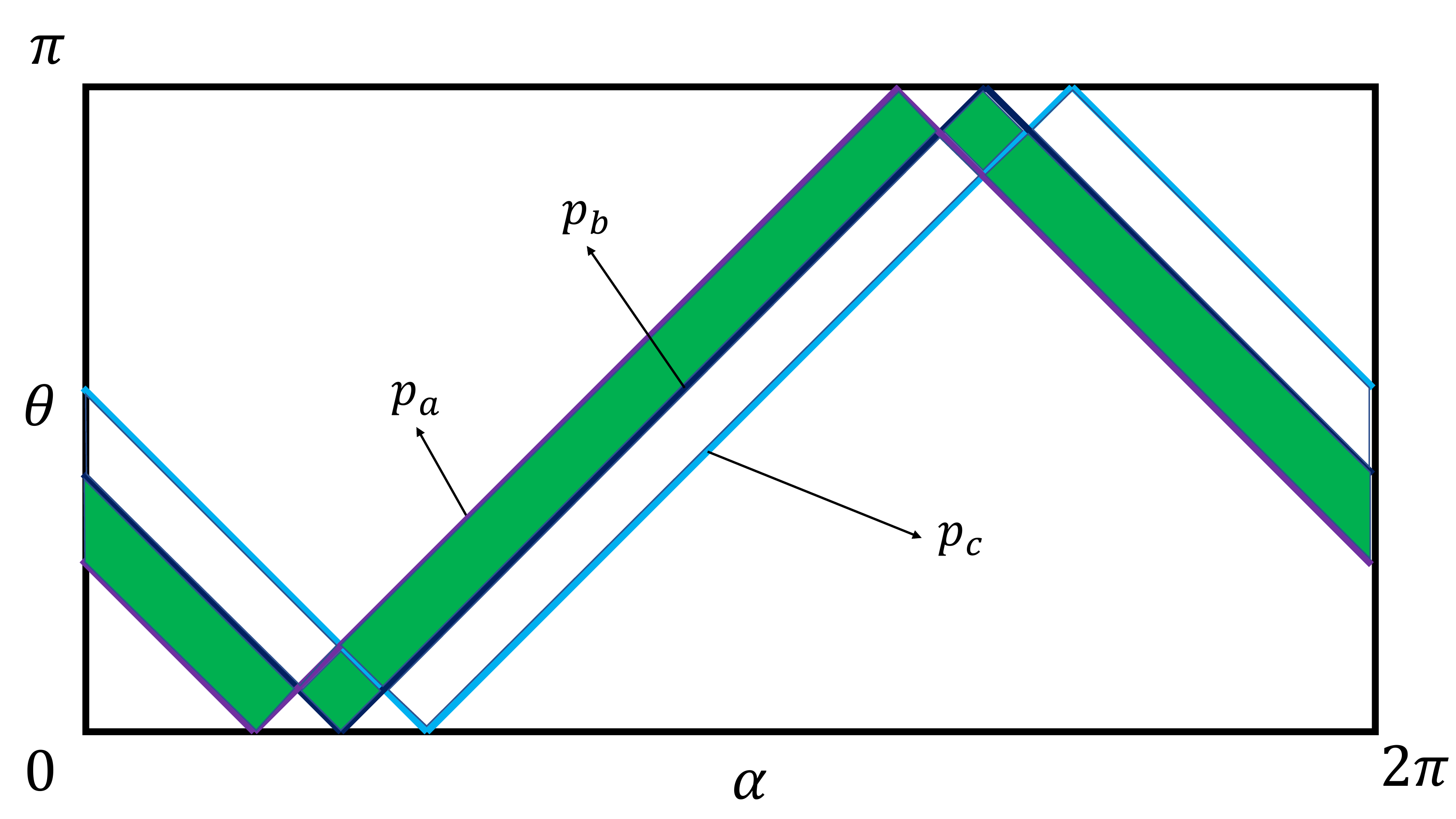} 
} 
\subfigure[] { \label{fig:b} 
\includegraphics[width=0.35\linewidth]{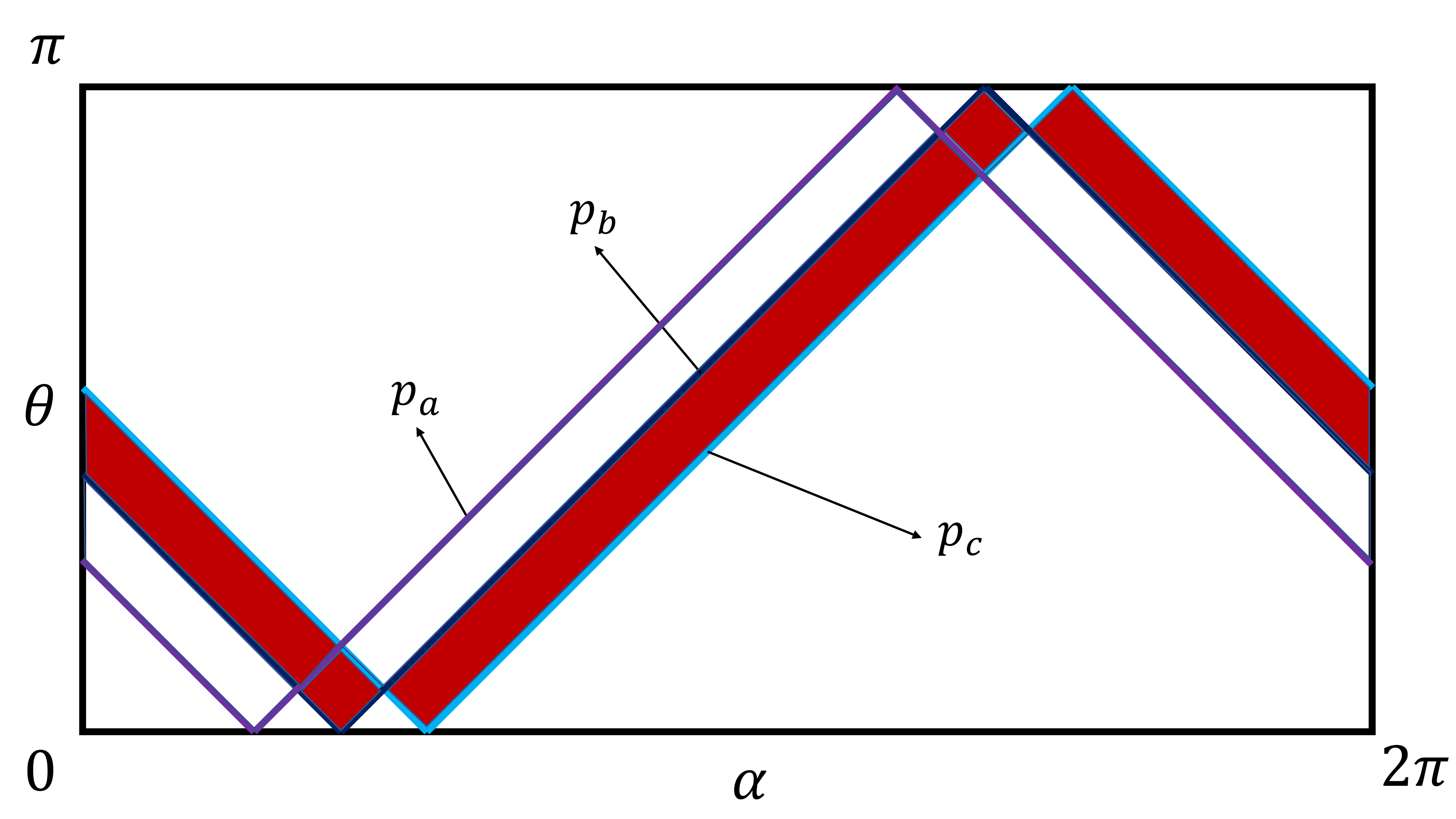} 
} 
\caption{The shaded regions represent the integral region. (a) The integral region of the interval $A$. (b) The integral region of the interval $B$.} 
\label{fig3} 
\end{figure} 

We are now interested in proving this inequality with the help of kinematic space. As mentioned previously, the area enclosed by the point curve of point $a$ and the point curve of point $b$ is shown in green, while the area enclosed by the point curve of point $b$ and the point curve of point $c$ is shown in red (as shown in Fig. \ref{fig3}).

The QES entropies for boundary intervals $A$, $B$ and $AB$ are given, respectively, by
\begin{eqnarray}
	S_{\text{QES}}(A)&=&\frac{1}{4} \int_{{p}_{a}\Delta {p}_{b}}\omega,\\
	S_{\text{QES}}(B)&=&\frac{1}{4} \int_{{p}_{b}\Delta{p}_{c}}\omega,\\
	S_{\text{QES}}(AB)&=&\frac{1}{4} \int_{{p}_{a}\Delta  {p}_{c}}\omega.
\end{eqnarray}

Due to the non-negativity of the integral measure and the area of the integral region satisfying ${p}_{a}\Delta  {p}_{b}+{p}_{b}\Delta  {p}_{c}\ge {p}_{a}\Delta  {p}_{c}$,  as shown in Fig. \ref{fig3}, we have
\begin{equation}
	\frac{1}{4} \int_{{p}_{a}\Delta  {p}_{b}}\omega+\frac{1}{4} \int_{{p}_{b}\Delta  {p}_{c}}\omega\ge \frac{1}{4} \int_{{p}_{a}\Delta  {p}_{c}}\omega.
\end{equation}
In other words
\begin{equation}
	S_{\text{QES}}(A)+S_{\text{QES}}(B)\ge S_{\text{QES}}(AB).
\end{equation}
More generally, we have 
$$\sum_{i=1}^{n} S_{\text{QES}}(A_{i} )\ge S_{\text{QES}}(\prod_{i=1}^{n} A_{i} ).$$

\subsubsection{Strong subadditivity}
\begin{figure}[t!]
	\centering
	\includegraphics[width=0.45\textwidth]{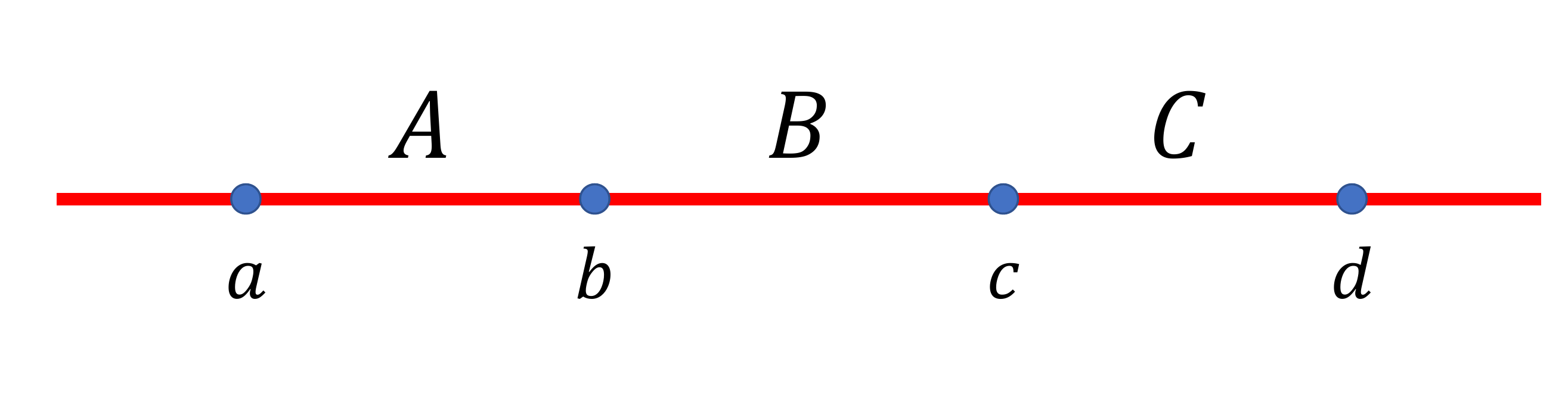}
	\caption{Subsystems on the boundary CFT, where $A$, $B$ and $C$ are three connected boundary intervals.}\label{3intervals}
\end{figure}

\begin{figure} \centering 
\subfigure[] { \label{fig:a} 
\includegraphics[width=0.35\linewidth]{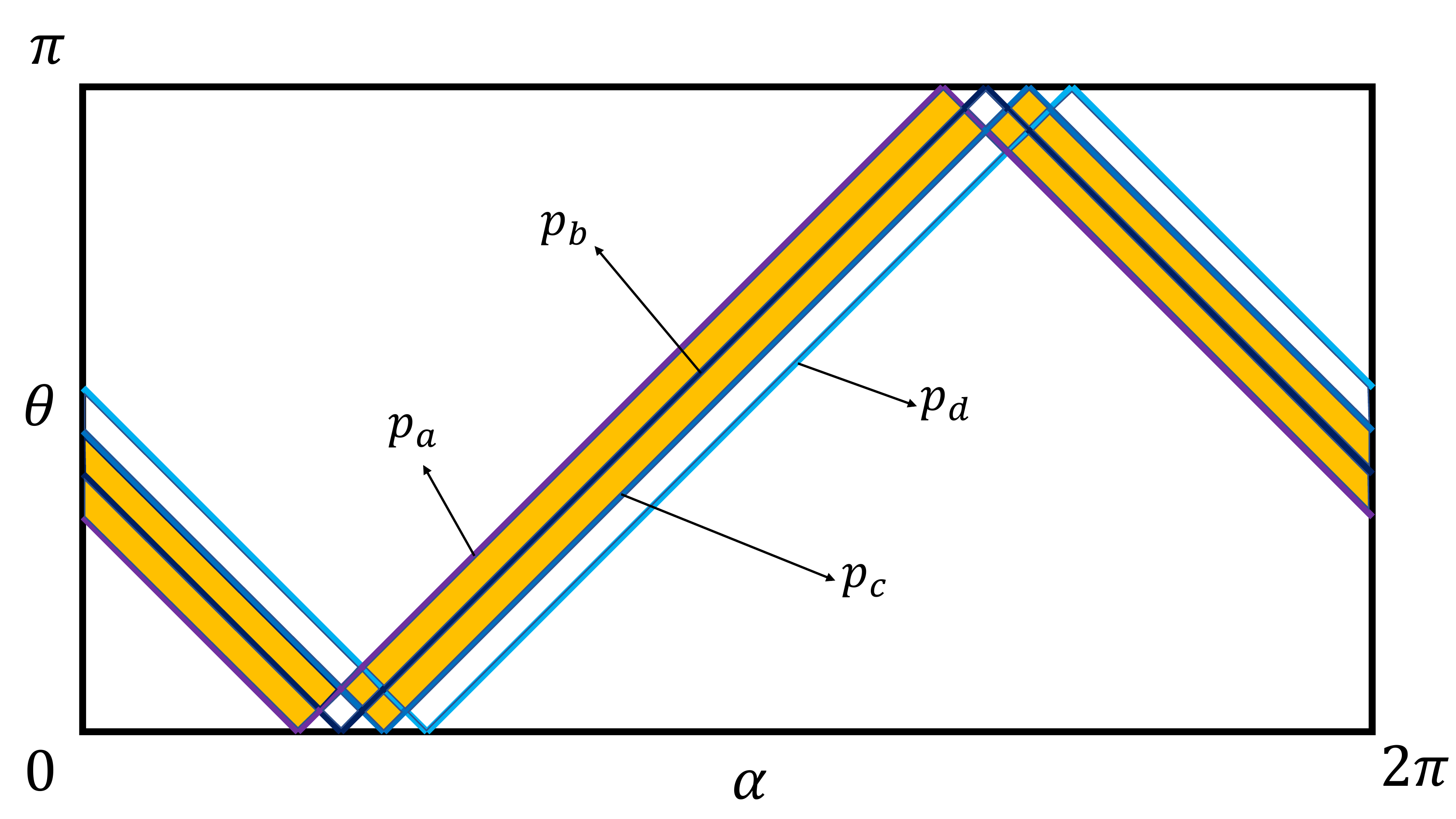} 
} 
\subfigure[] { \label{fig:b} 
\includegraphics[width=0.35\linewidth]{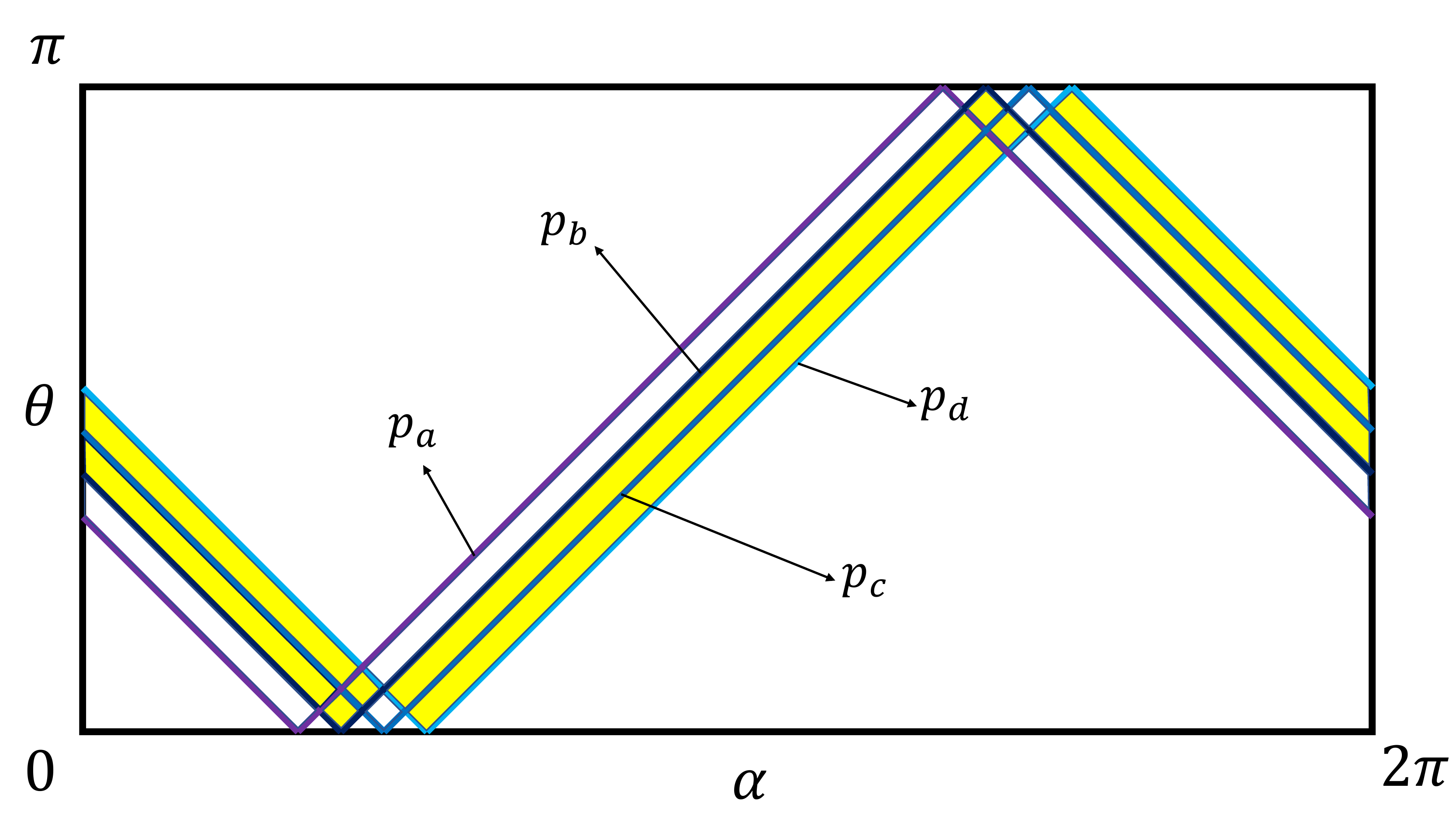} 
} 
\caption{The shaded regions represent the integral region. (a) The integral region of the interval $AB$. (b) The integral region of the interval $BC$.} 
\label{fig6} 
\end{figure}

\begin{figure} \centering 
\subfigure[] { \label{fig:a} 
\includegraphics[width=0.35\linewidth]{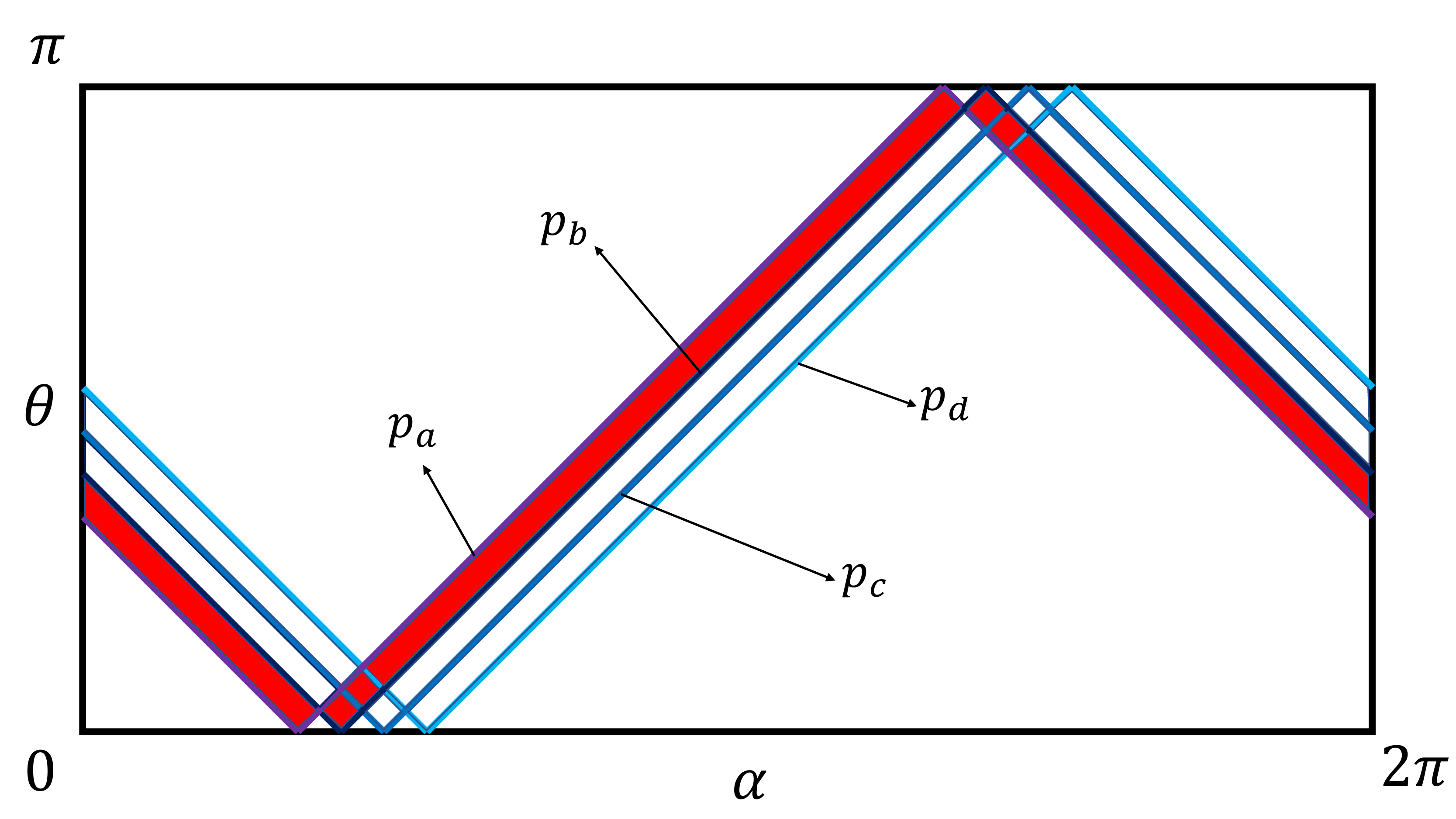} 
} 
\subfigure[] { \label{fig:b} 
\includegraphics[width=0.35\linewidth]{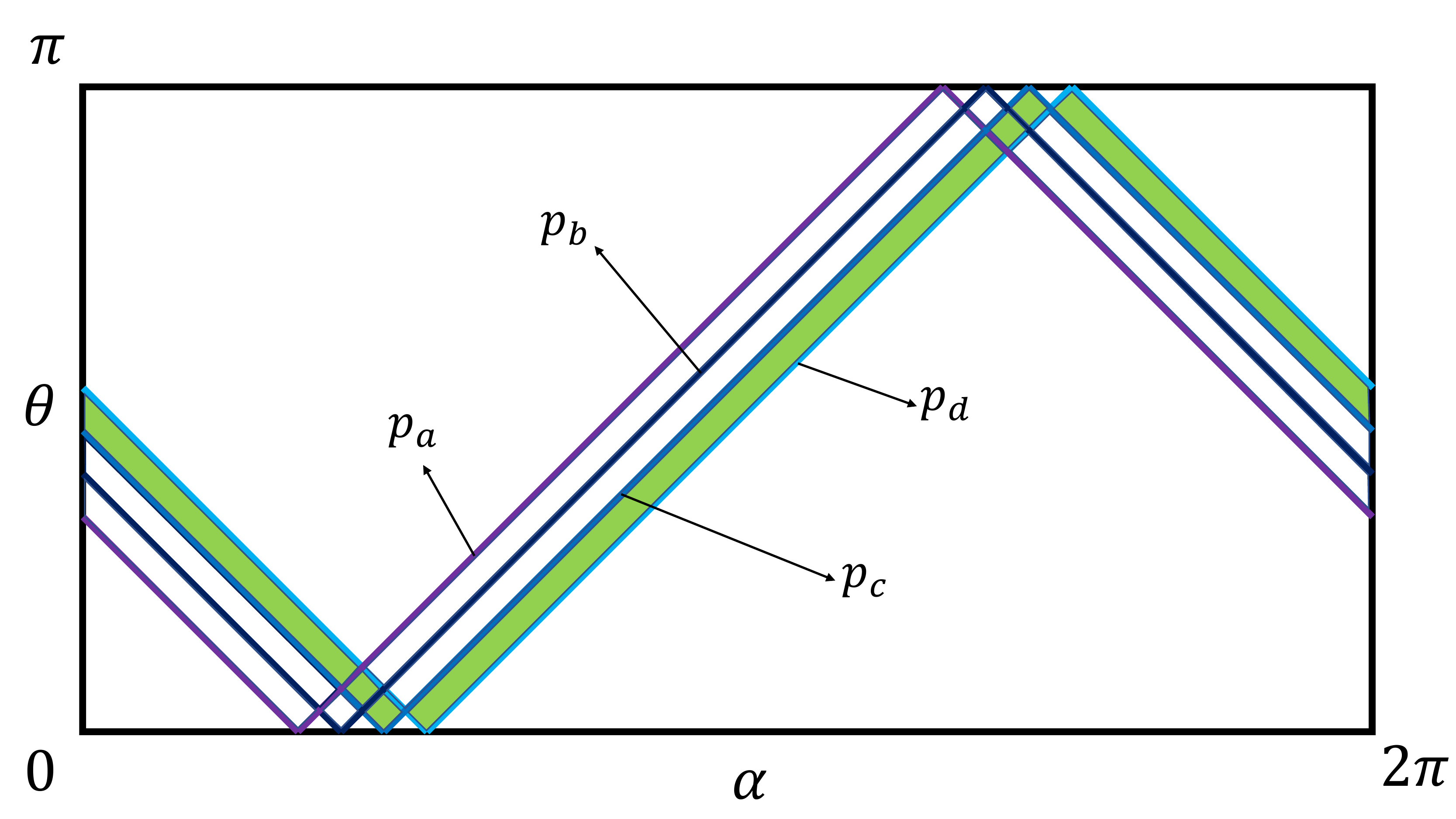} 
} 
\caption{The shaded regions represent the integral region. (a) The integral region of the interval $A$. (b) The integral region of the interval $C$.} 
\label{fig7} 
\end{figure}

The strong subadditivity states that for three non-overlapping boundary regions $A$, $B$ and $C$ (as shown in Fig. \ref{3intervals}), we have the following inequality for the associated entanglement entropies
\begin{equation}
S_{\text{QES}}(AB)+S_{\text{QES}}(BC)\ge S_{\text{QES}}(A)+S_{\text{QES}}(C).
\end{equation}
Let $A$, $B$ and $C$ be three joint boundary intervals, $a$ and $b$ are the endpoints of interval $A$, $b$ and $c$ are the endpoints of interval $B$, and $c$ and $d$ are the endpoints of interval $C$, as shown in Fig. \ref{3intervals}. The point curve of point a is $p_a$, the point curve of point $b$ is $p_b$, the point curve of point $c$ is $p_c$, and the point curve of point $d$ is $p_d$. The entanglement entropy of boundary intervals $A$, $C$, $AB$ and $BC$ are given, respectively, by
\begin{eqnarray}
	S_{\text{QES}}(A)&=& \frac{1}{4} \int_{{p}_{a}\Delta  {p}_{b}}\omega,\\
	S_{\text{QES}}(C)&=& \frac{1}{4} \int_{{p}_{c}\Delta  {p}_{d}}\omega,\\
	S_{\text{QES}}(AB)&=& \frac{1}{4} \int_{{p}_{a}\Delta  {p}_{c}}\omega,\\
	S_{\text{QES}}(BC)&=& \frac{1}{4} \int_{{p}_{b}\Delta  {p}_{d}}\omega.
\end{eqnarray}
As can be seen from the kinematic space (as shown in the Fig. \ref{fig6} and Fig. \ref{fig7}), due to the non-negativity of the integral measure, we can directly derive:
\begin{equation}
\frac{1}{4} \int_{{p}_{a}\Delta  {p}_{c}}\omega+\frac{1}{4} \int_{{p}_{b}\Delta  {p}_{d}}\omega\ge \frac{1}{4} \int_{{p}_{a}\Delta  {p}_{b}}\omega+\frac{1}{4} \int_{{p}_{c}\Delta  {p}_{d}}\omega.
\end{equation}
This leads to the strong subadditivity:
\begin{equation}
S_{\text{QES}}(AB)+S_{\text{QES}}(BC)\ge S_{\text{QES}}(A)+S_{\text{QES}}(C).	
\end{equation}

\subsubsection{Subadditivity of conditional entropy}
Given two disjoint boundary regions $A$ and $B$, the conditional entropy is defined as follows:
 \begin{equation}
S_{\text{QES}}(A\mid B)=S_{\text{QES}}(AB)-S_{\text{QES}}(B).
 \end{equation}
Then the subadditivity of conditional entropy states that for three non-overlapping boundary regions $A$, $B$ and $C$, the associated conditional entanglement entropies should obey the following relationship:
 \begin{equation}
 S_{\text{QES}}(A\mid B)+S_{\text{QES}}(C\mid B)\ge S_{\text{QES}}(AC\mid B).
 \end{equation}
 
 \begin{figure} \centering 
\subfigure[] { \label{fig:a} 
\includegraphics[width=0.35\linewidth]{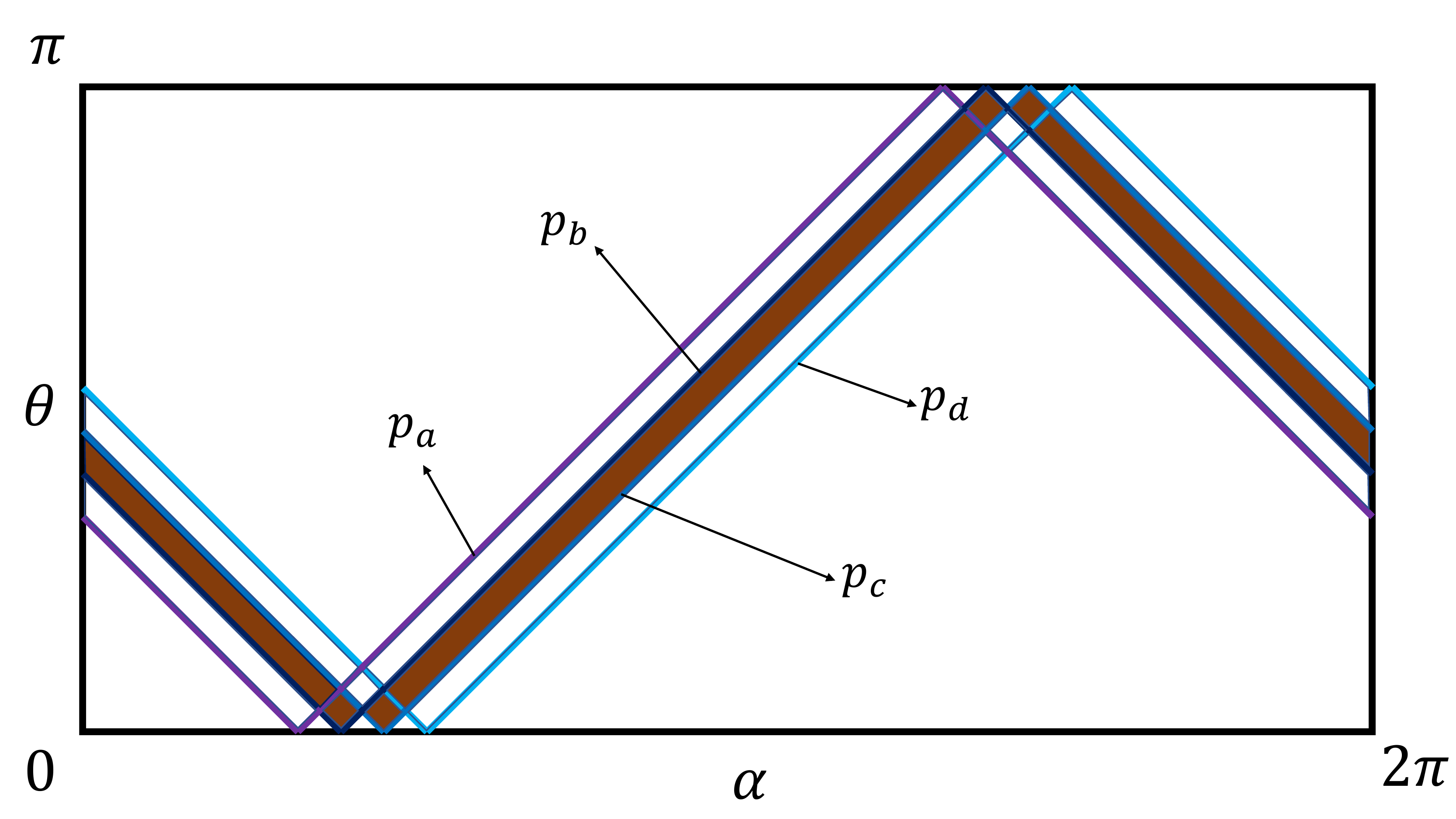} 
} 
\subfigure[] { \label{fig:b} 
\includegraphics[width=0.35\linewidth]{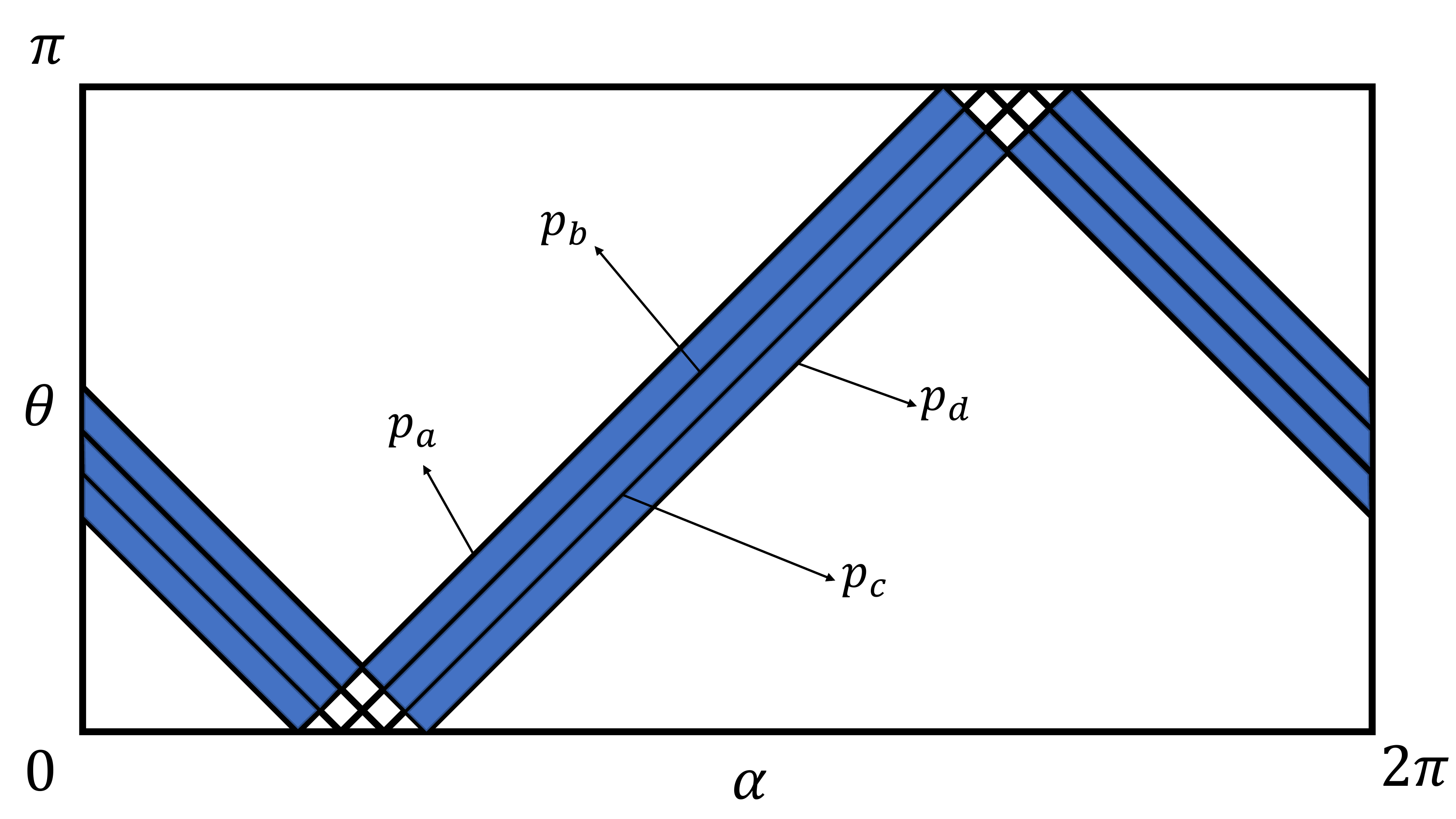} 
} 
\caption{The shaded regions represent the integral region. (a) The integral region of the interval $B$. (b) The integral region of the interval $ABC$.} 
\label{fig8} 
\end{figure}

According to the definition, the conditional entropy can be obtained by integrating over the corresponding regions in the kinematic space
\begin{equation}
	S_{\text{QES}}(A\mid B)= \frac{1}{4} \int_{{p}_{a}\Delta  {p}_{c}}\omega-\frac{1}{4} \int_{{p}_{b}\Delta  {p}_{c}}\omega,
\end{equation}
\begin{equation}
	S_{\text{QES}}(C\mid B)= \frac{1}{4} \int_{{p}_{b}\Delta  {p}_{d}}\omega-\frac{1}{4} \int_{{p}_{b}\Delta  {p}_{c}}\omega,
\end{equation}
\begin{equation}
	S_{\text{QES}}(AC\mid B)= \frac{1}{4} \int_{{p}_{a}\Delta  {p}_{d}}\omega-\frac{1}{4} \int_{{p}_{b}\Delta  {p}_{c}}\omega.
\end{equation}
Then it is directly to show by observing the integral area in Fig. \ref{fig6} and Fig. \ref{fig8}
\begin{equation}
S_{\text{QES}}(A\mid B)+S_{\text{QES}}(C\mid B)\ge S_{\text{QES}}(AC\mid B).	
\end{equation}

\begin{figure}
	\centering
	\includegraphics[width=0.45\textwidth]{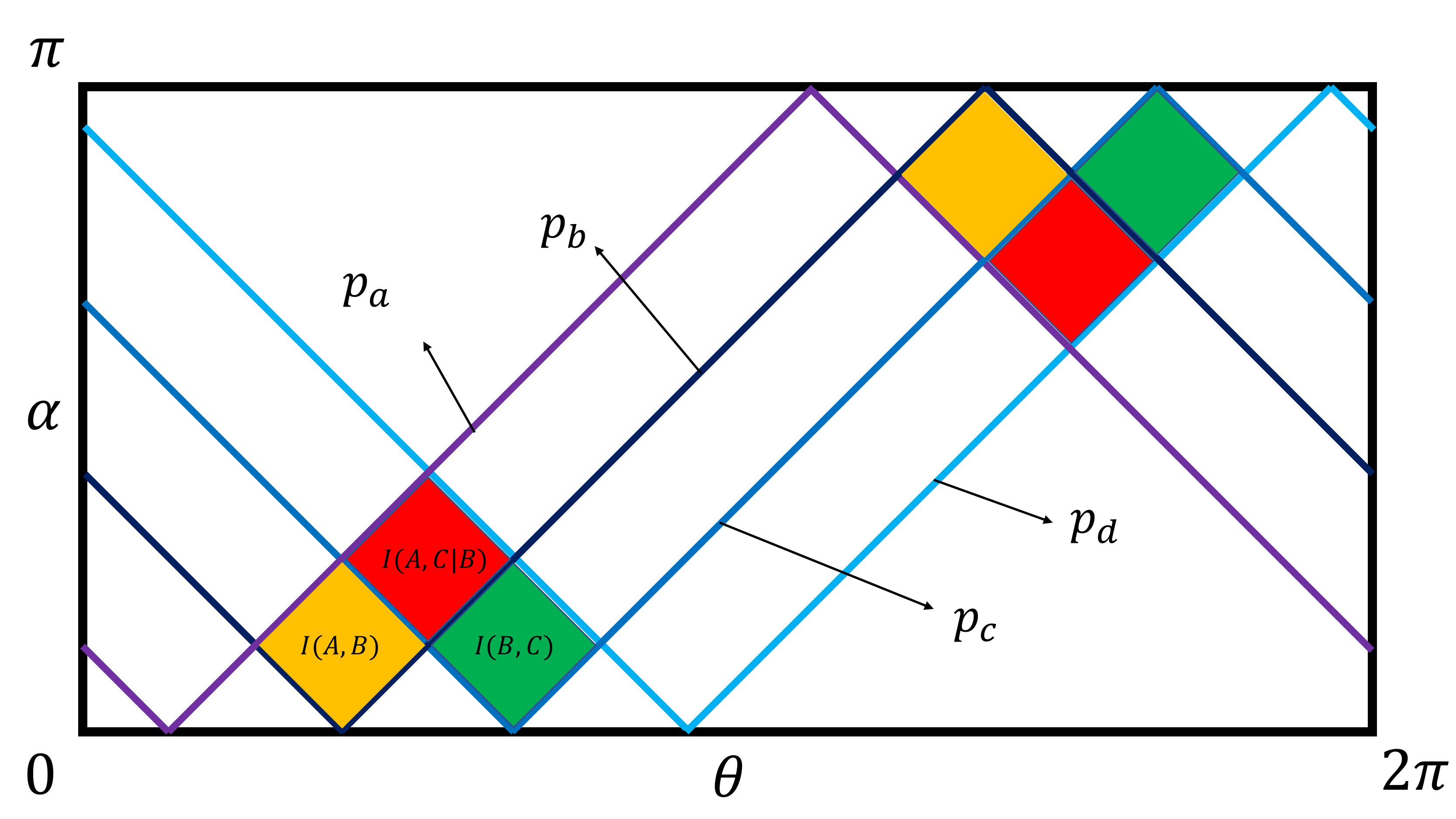}
	\caption{The yellow area is used to calculate the mutual information between $A$ and $B$, the green area is used to calculate the mutual information between $B$ and $C$, and the red area is used to calculate the conditional mutual information.}
	\label{mi}
\end{figure}

Another advantage of kinematic space is that the mutual information between several entangled regions can be intuitively understood by examining the kinematic space, as depicted in Fig. \ref{mi}. In other words, the mutual information and conditional mutual information between multiple regions correspond to specific regions within the kinematic space. By integrating these regions with appropriate integral measures, we can obtain the mutual information and conditional mutual information between the entanglement intervals.

\subsection{Proof of QES entropy inequalities in the EOW description} 
In the previous subsection, we demonstrated how QES entropy inequalities can be proven in the AdS description using the concept of kinematic space. However, this proof relies heavily on the assumption of strong subadditivity \eqref{strongsubadd} in order to ensure the non-negativity of the integral measure. In this subsection, we will present an alternative proof in the context of the EOW description (double holography viewpoint) which does not require this assumption. Furthermore, we will show that not only is it unnecessary to assume strong subadditivity \eqref{strongsubadd} beforehand, but strong subadditivity itself can be naturally derived within this framework.

\subsubsection{Subadditivity}
From viewpoint of the double holography, a $(d+1)$-dimensional bulk archors an EOW membrane. The entanglement entropy of region $R$ in BCFT can be calculated using the RT surface. The two endpoints of the RT surfaces are located, respectively, at the boundary of region $R$ and the boundary of the island corresponding to $R$, as shown in Fig. \ref{fig:c}. When the two endpoints $(u, v)$ of region $R$ are determined, the corresponding holographic entanglement entropy is also determined. Two intersection points form a quantum extreme island on the membrane. The two endpoints of the quantum extreme island are denoted as $(I_u, I_v)$. The length of a geodesic connecting two points $u$ and $I_u$ can be calculated using the region enclosed by the point curves of these two points in the kinematic space. Thus, after selecting a time slice, holographic entanglement entropy of $R$ region can be expressed as
\begin{equation}
	S_{\text{QES}}(R)=\frac{A^{(d)}\left(\partial I_{R}\right)}{4 G^{(d)}}+\frac{1}{4} \int_{p_{u} \Delta p_{I_{u}}} \omega+\frac{1}{4} \int_{p_{v} \Delta p_{I_{v}}} \omega.
\end{equation}

Consider two joint subsystems, $A$ and $B$, on the boundary of the CFT. Taking a fixed time slice, the RT surface is a geodesic on this time slice. The geodesic starts at an endpoint, $a_1$, of region $A$ and ends at the boundary, $I_{a_1}$, of the quantum extreme island corresponding to $A$. The length of this geodesic can be calculated using the Crofton formula. Therefore, as shown in Fig. \ref{st}, the entropy of region $A$ can be expressed as follows:
\begin{figure}
	\centering
	\includegraphics[width=0.45\textwidth]{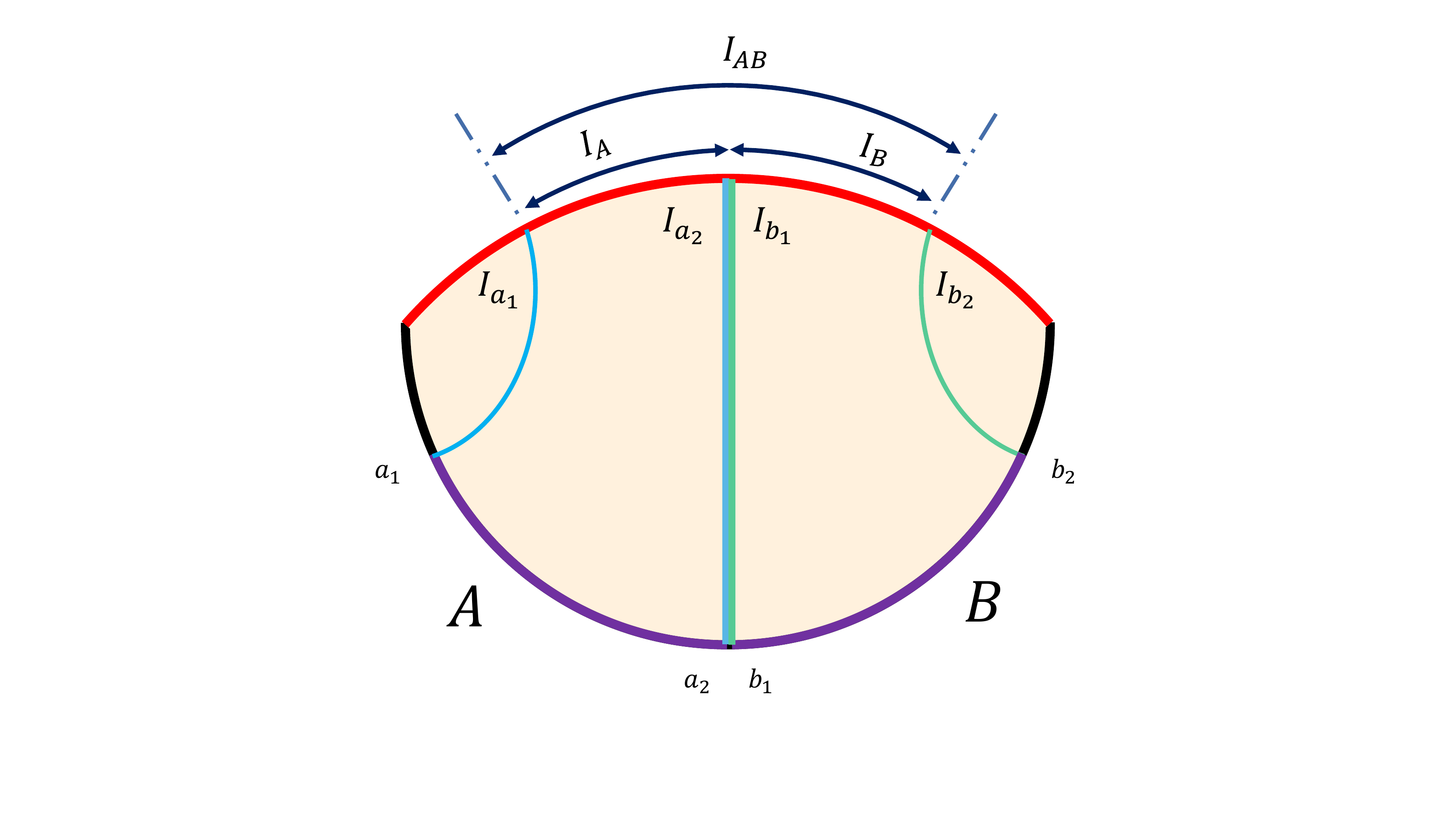}
	\includegraphics[width=0.45\textwidth]{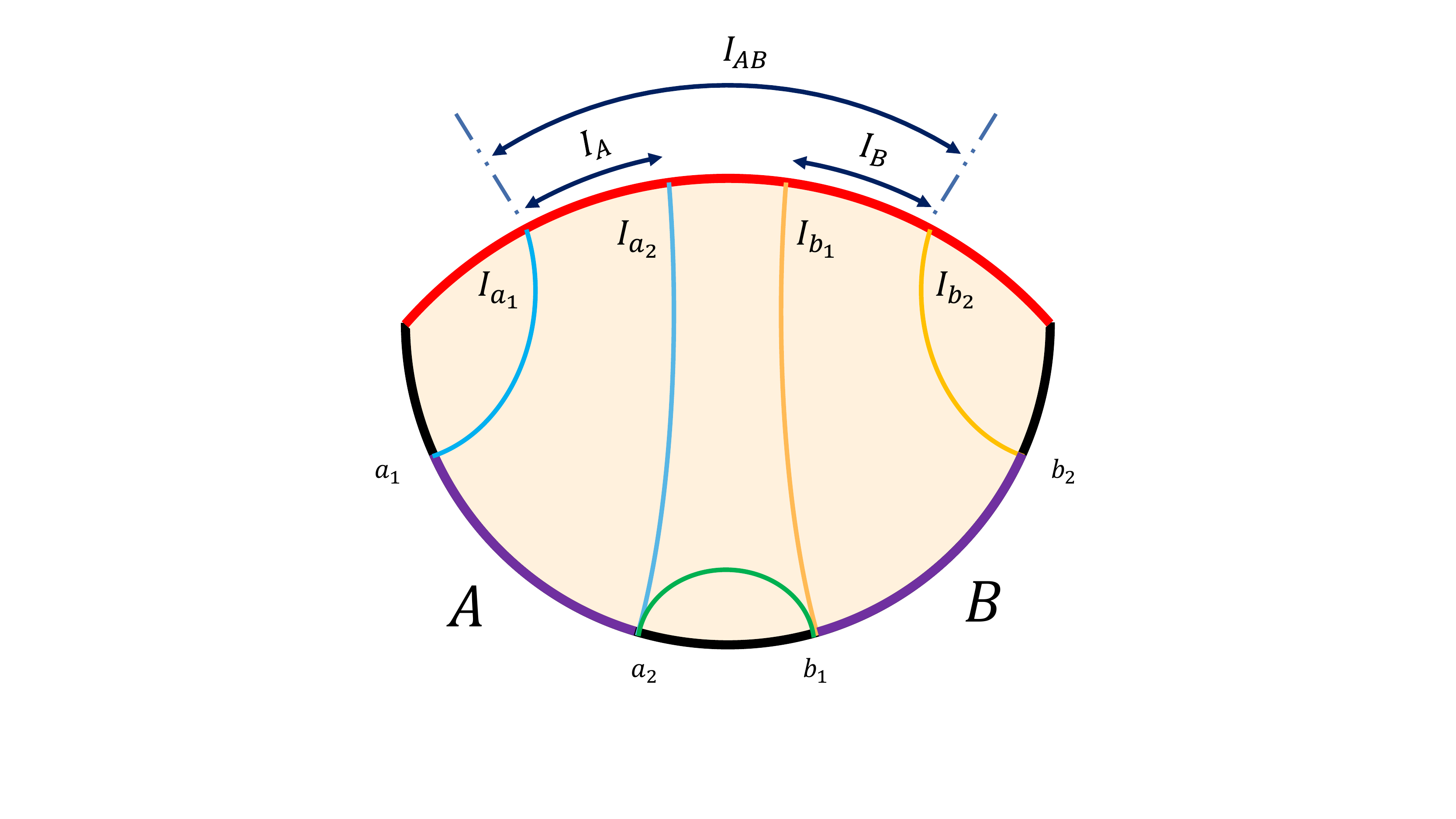}
	\caption{Left: The subsystems $A$ and $B$ are connected. The blue curves correspond to the RT surface of interval $A$. The green curves denote to the RT surface of interval $B$. Right: The subsystems $A$ and $B$ are disconnected.  The blue curves correspond to the RT surface of interval $A$. The orange curves correspond to the RT surface of interval $B$.}
	\label{st}
\end{figure}
\begin{equation}\label{sqesa}
	S_{\text{QES}}(A)=\frac{A^{(d)}\left(\partial I_{A}\right)}{4 G^{(d)}}+\frac{1}{4} \int_{p_{a_1} \Delta p_{I_{a_1}}} \omega+\frac{1}{4} \int_{p_{a_2} \Delta p_{I_{a_2}}} \omega.
\end{equation}
The integral measure here is defined as
\begin{equation}
	\omega = \frac{\partial^2 l(u,v)}{\partial u\, \partial v} \, dudv.\
\end{equation}
Note that $l(u, v)$ here is the length of the geodesic connecting two points on the boundary. Therefore the non-negativity of the measure $\omega$ is automatically satisfied.

Similarly, the entropy of region $B$ can be expressed as
\begin{equation}\label{sqesb}
	S_{\text{QES}}(B)=\frac{A^{(d)}\left(\partial I_{B}\right)}{4 G^{(d)}}+\frac{1}{4} \int_{p_{b_1} \Delta p_{I_{b_1}}} \omega+\frac{1}{4} \int_{p_{b_2} \Delta p_{I_{b_2}}} \omega.
\end{equation}

It is worth mentioning that the contribution of the holographic entanglement entropy from bulk corresponds to the integral region in the kinematic space (take the geodesic connecting two points $a_1$ and $I_ {a_1}$ in the figure as an example) as shown on the left panel of Fig. \ref{st}.

On the other hand, it can be shown that the entropy of the $AB$ region follows
\begin{equation}\label{sqesab}
	S_{\text{QES}}(AB)=\frac{A^{(d)}\left(I_{a_1}\right)}{4 G^{(d)}}+\frac{1}{4} \int_{p_{a_1} \Delta p_{I_{a_1}}} \omega+\frac{A^{(d)}\left(I_{b_2}\right)}{4 G^{(d)}}+\frac{1}{4} \int_{p_{b_2} \Delta p_{I_{b_2}}} \omega.
\end{equation}
Combining Eqs. \eqref{sqesa}, \eqref{sqesb} and \eqref{sqesab} we finally find that subadditivity holds
\begin{equation}
	S_{\text{QES}}(A)+S_{\text{QES}}(B)\ge S_{\text{QES}}(AB).
\end{equation}

For two disjoint intervals $A$ and $B$  (as shown on the right side of Fig. \ref{st}), let's prove the subadditivity inequality in this case. From the description in Fig. \ref{st}, it can be seen that the entanglement entropy of intervals $A$  and $B$  is given by
\begin{equation}\label{sqa}
	S_{\text{QES}}(A)=\frac{A^{(d)}\left(\partial I_{A}\right)}{4 G^{(d)}}+\frac{1}{4} \int_{p_{a_1} \Delta p_{I_{a_1}}} \omega+\frac{1}{4} \int_{p_{a_2} \Delta p_{I_{a_2}}} \omega,
\end{equation}
\begin{equation}\label{sqb}
	S_{\text{QES}}(B)=\frac{A^{(d)}\left(\partial I_{B}\right)}{4 G^{(d)}}+\frac{1}{4} \int_{p_{b_1} \Delta p_{I_{b_1}}} \omega+\frac{1}{4} \int_{p_{b_2} \Delta p_{I_{b_2}}} \omega.
\end{equation}

From a holographic perspective, the entanglement entropy of a disjoint interval $AB$  can be determined by considering two possible cases and selecting the minimum entropy value. This is expressed as:
\begin{equation}\label{sab}
S_{\text{QES}} (AB) = \min(S_{a_{1}a_{2}}+S_{b_{1}b_{2}}, S_{a_{1}b_{2}}+S_{a_{2}b_{1}}),
\end{equation}
Here, $S_{a_{1}a_{2}}$ represents the entropy of interval $a_{1}a_{2}$, and the other similar symbols in the equation carry the same meaning. 

When the entanglement entropy of the disjoint interval $AB$  is determined by selecting the case where the sum of entropies between intervals $a_{1}a_{2}$ and $b_{1}b_{2}$ is the minimum in \eqref{sab}, it is straightforward to show
\begin{equation}
	S_{\text{QES}}(A)+S_{\text{QES}}(B)= S_{\text{QES}}(AB).
\end{equation}

Instead, as the sum of the entropies of interval $a_{1}b_{2}$ and interval $a_{2}b_{1}$ is minimum of \eqref{sab}, i.e., $S_{\text{QES}} (AB) = S_{a_{1}b_{2}}+S_{a_{2}b_{1}}$. Then it is not difficult to show
\begin{equation}
	\frac{A^{(d)}\left( I_{a_1}\right)}{4 G^{(d)}}+\frac{1}{4} \int_{p_{a_1} \Delta p_{I_{a_1}}} \omega+\frac{A^{(d)}\left( I_{b_2}\right)}{4 G^{(d)}}+\frac{1}{4} \int_{p_{b_2} \Delta p_{I_{b_2}}} \omega \ge S_{a_1b_2},
\end{equation}
\begin{equation}
	\frac{A^{(d)}\left(I_{a_2}\right)}{4 G^{(d)}}+\frac{1}{4} \int_{p_{a_2} \Delta p_{I_{a_2}}} \omega+\frac{A^{(d)}\left(I_{b_1}\right)}{4 G^{(d)}}+\frac{1}{4} \int_{p_{b_1} \Delta p_{I_{b_1}}} \omega \ge S_{a_2b_1}.
\end{equation}
Combining \eqref{sqa} and \eqref{sqb} we find that the subadditivity inequality for disjoint intervals $A$  and $B$  holds
\begin{equation}
	S_{\text{QES}}(A)+S_{\text{QES}}(B)\ge S_{\text{QES}}(AB).
\end{equation}
\subsubsection{Strong subadditivity}
Let us now turn to prove the strong subadditivity inequality within the context of double holography. The strong subadditivity inequality is expressed as follows: 
\begin{equation}
	S_{\text{QES}}(AB)+S_{\text{QES}}(BC)\ge S_{\text{QES}}(A)+S_{\text{QES}}(C).
\end{equation}
Similarly, there are two types of entanglement intervals: one is  that the three intervals $A$ , $B$, and $C$ are connected, the other is that the three intervals $A$ , $B$, and $C$ are disconnected. 

\begin{figure}
	\centering
	\includegraphics[width=0.45\textwidth]{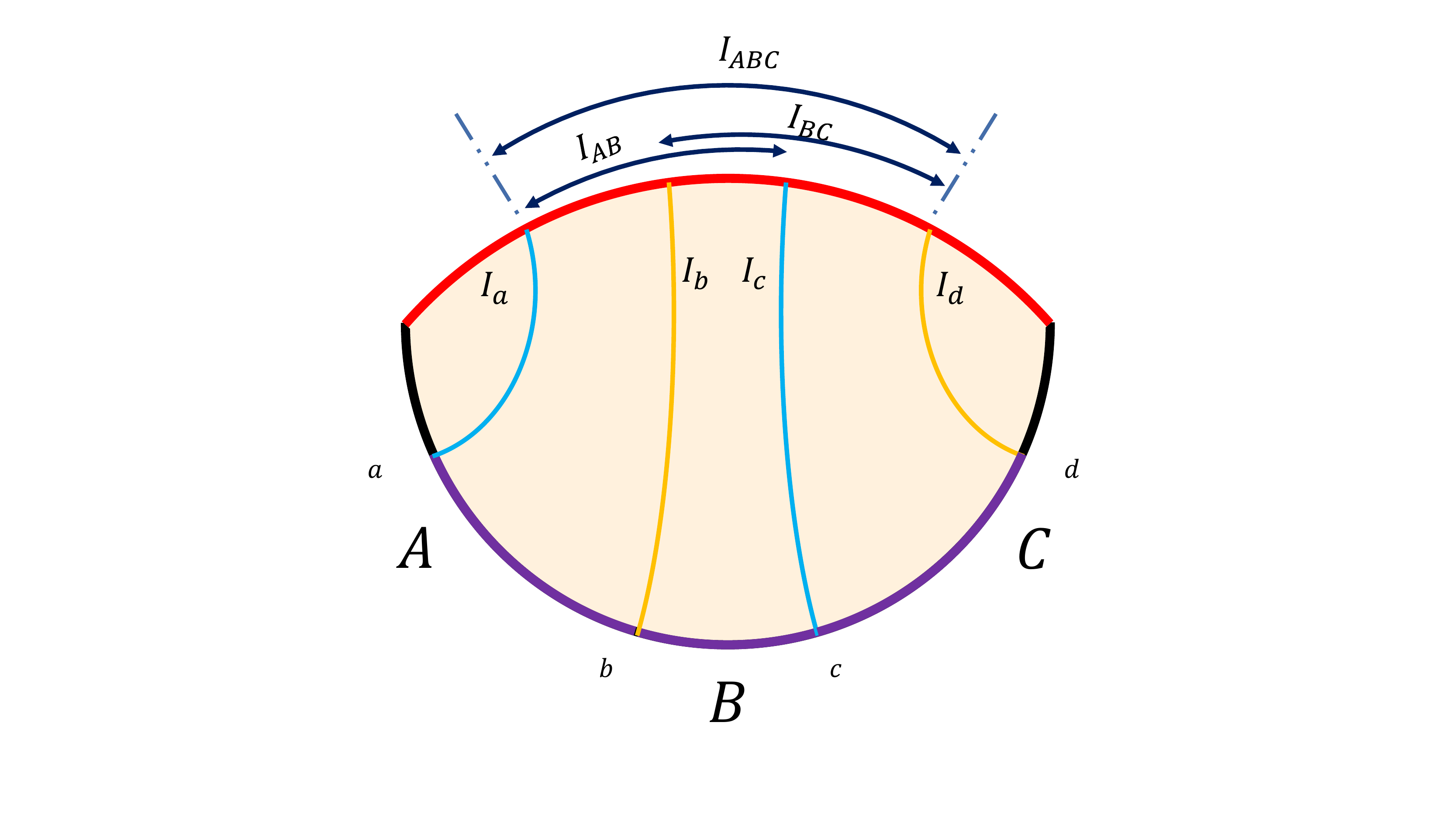}
	\includegraphics[width=0.45\textwidth]{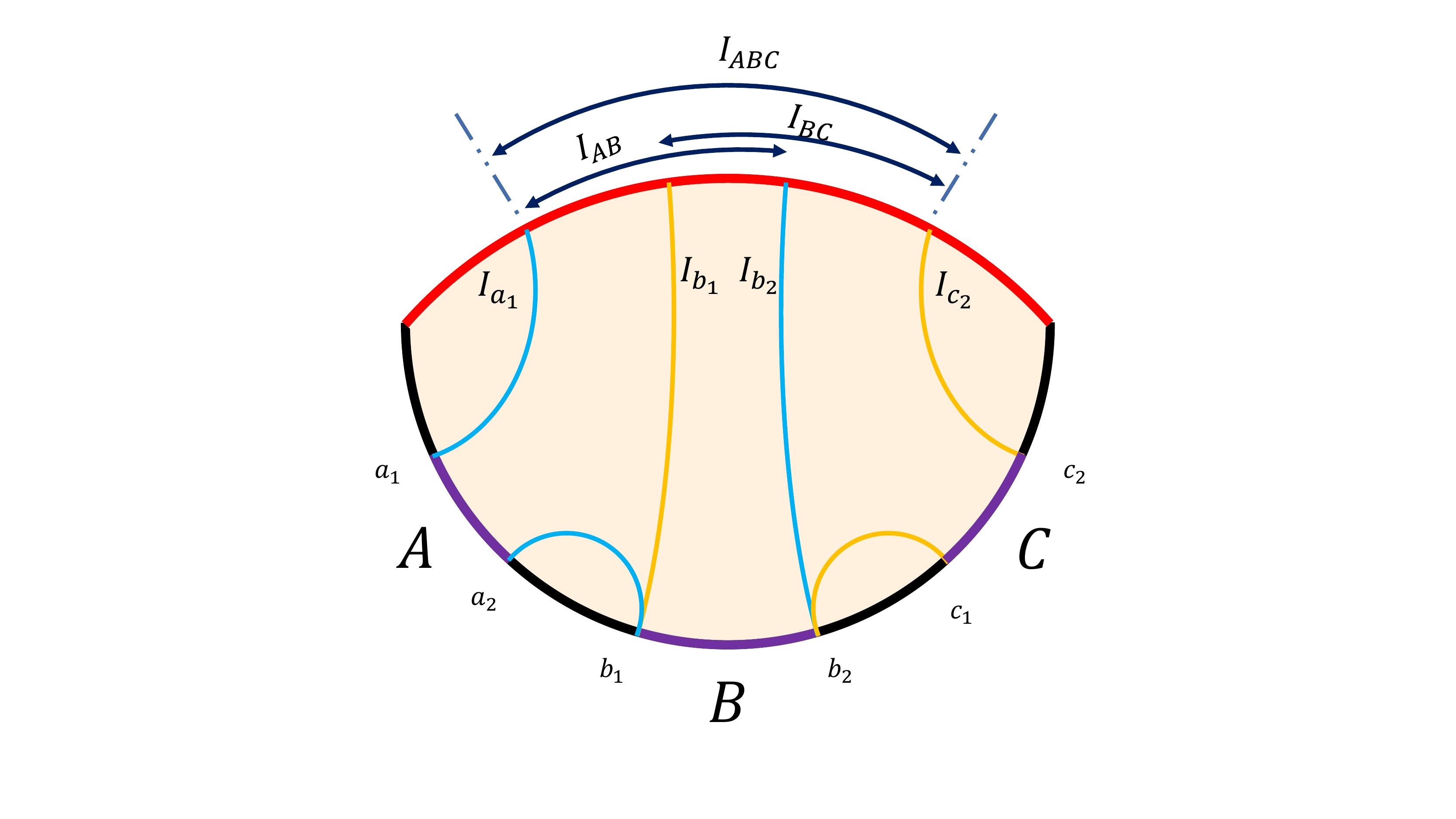}
	\caption{Left: The RT surfaces for interval $AB$ (the blue curves) and $BC$ (the orange curves) as they are connected. Right: The RT surfaces for interval $AB$ (the blue curves) and $BC$ (the orange curves)  as they are disconnected.}
	\label{ssa}
\end{figure}

Let us consider the former case firstly. As shown in Fig. \ref{ssa}, the entanglement entropy of intervals $AB$ and $BC$ is

\begin{equation}
		 S_{\text{QES}}(AB)=\frac{A^{(d)}\left(I_{a}\right)}{4 G^{(d)}}+\frac{1}{4} \int_{p_{a} \Delta p_{I_{a}}} \omega+\frac{A^{(d)}\left(I_{c}\right)}{4 G^{(d)}}+\frac{1}{4} \int_{p_{c} \Delta p_{I_{c}}} \omega, 
\end{equation}
\begin{equation}
	S_{\text{QES}}(BC)=\frac{A^{(d)}\left(I_{b}\right)}{4 G^{(d)}}+\frac{1}{4} \int_{p_{b} \Delta p_{I_{b}}} \omega+\frac{A^{(d)}\left(I_{d}\right)}{4 G^{(d)}}+\frac{1}{4} \int_{p_{d} \Delta p_{I_{d}}} \omega.
\end{equation}
By recombining the items in the above equation, we can obtain
\begin{equation}
	\frac{A^{(d)}\left(I_{a}\right)}{4 G^{(d)}}+\frac{1}{4} \int_{p_{a} \Delta p_{I_{a}}} \omega+\frac{A^{(d)}\left(I_{b}\right)}{4 G^{(d)}}+\frac{1}{4} \int_{p_{b} \Delta p_{I_{b}}} \omega\ge S_{\text{QES}}(A),
\end{equation}
\begin{equation}
	\frac{A^{(d)}\left(I_{c}\right)}{4 G^{(d)}}+\frac{1}{4} \int_{p_{c} \Delta p_{I_{c}}} \omega+\frac{A^{(d)}\left(I_{d}\right)}{4 G^{(d)}}+\frac{1}{4} \int_{p_{d} \Delta p_{I_{d}}} \omega\ge S_{\text{QES}}(C).
\end{equation}
This proves the existence of strong subadditivity.

Then let us consider the strongly subadditive inequalities for three disjoint intervals, as shown in Fig. \ref{ssa}. The entropy of subsystems $AB$ and $BC$ is calculated using the RT surface and the brane contributions of the boundaries of their associated islands on the brane, $I_{AB}$ and $I_{BC}$, respectively. The entanglement entropy of subsystems $AB$ and $BC$ can be expressed as:
\begin{equation}
S_{\text{QES}}(AB)=\frac{A^{(d)}\left(\partial I_{AB}\right)}{4 G^{(d)}}+\frac{1}{4} \int_{p_{a_1} \Delta p_{I_{a_1}}} \omega+\frac{1}{4} \int_{p_{a_2} \Delta p_{b_1}} \omega+\frac{1}{4} \int_{p_{b_2} \Delta p_{I_{b_2}}} \omega,
\label{S(AB)}
\end{equation}

\begin{equation}
S_{\text{QES}}(BC)=\frac{A^{(d)}\left(\partial I_{BC}\right)}{4 G^{(d)}}+\frac{1}{4} \int_{p_{b_1} \Delta p_{I_{b_1}}} \omega+\frac{1}{4} \int_{p_{b_2} \Delta p_{c_1}} \omega+\frac{1}{4} \int_{p_{c_2} \Delta p_{I_{c_2}}} \omega.
\label{S(BC)}
\end{equation}

By reassembling the terms on the right-hand side of the equations, we observe that this recombination corresponds to subsystems $A$ and $C$. The reconstructed terms are greater than or equal to the entropy of their corresponding subsystems:
\begin{equation}
\frac{A^{(d)}\left(I_{a_1}\right)}{4 G^{(d)}}+\frac{A^{(d)}\left(I_{b_1}\right)}{4 G^{(d)}}+\frac{1}{4} \int_{p_{a_1} \Delta p_{I_{a_1}}} \omega+\frac{1}{4} \int_{p_{a_2} \Delta p_{b_1}} \omega+\frac{1}{4} \int_{p_{b_1} \Delta p_{I_{b_1}}} \omega \ge S_{\text{QES}}(A),
\end{equation}
\begin{equation}
\frac{A^{(d)}\left(I_{b_2}\right)}{4 G^{(d)}}+\frac{A^{(d)}\left(I_{c_2}\right)}{4 G^{(d)}}+\frac{1}{4} \int_{p_{b_2} \Delta p_{I_{b_2}}} \omega+\frac{1}{4} \int_{p_{b_2} \Delta p_{c_1}} \omega+\frac{1}{4} \int_{p_{c_2} \Delta p_{I_{c_2}}} \omega \ge S_{\text{QES}}(C).
\end{equation}
By combining these four equations, it can be shown that any three subsystems $A$ , $B$ and $C$ obey the strong subadditivity inequalities:
\begin{equation}
	S_{\text{QES}}(AB)+S_{\text{QES}}(BC)\ge S_{\text{QES}}(A)+S_{\text{QES}}(C).
\end{equation}
This ends the proof.

In addition, using the similar approach, one can prove that the following form of strongly subadditivity inequality is also valid
\begin{equation}
	S_{\text{QES}}(AB)+S_{\text{QES}}(BC)\ge S_{\text{QES}}(B)+S_{\text{QES}}(ABC).
\end{equation}
As shown on the right side of Fig. \ref{ssa}, the entanglement entropy of intervals $AB$ and $BC$ is the same as equations \eqref{S(AB)} and \eqref{S(BC)}, so we can obtain
\begin{align}
	&\frac{A^{(d)}\left(I_{a_1}\right)}{4 G^{(d)}}+\frac{1}{4} \int_{p_{a_1} \Delta p_{I_{a_1}}} \omega+\frac{A^{(d)}\left(I_{c_2}\right)}{4 G^{(d)}}+\frac{1}{4} \int_{p_{c_2} \Delta p_{I_{c_2}}} \omega+\frac{1}{4} \int_{p_{a_2} \Delta p_{b_1}} \omega+\frac{1}{4} \int_{p_{b_2} \Delta p_{c_1}} \omega \notag
	\\
	&\ge S_{QES}(ABC),
\end{align}
\begin{equation}
	\frac{A^{(d)}\left(I_{b_2}\right)}{4 G^{(d)}}+\frac{1}{4} \int_{p_{b_2} \Delta p_{I_{b_2}}} \omega+\frac{A^{(d)}\left(I_{b_1}\right)}{4 G^{(d)}}+\frac{1}{4} \int_{p_{b_1} \Delta p_{I_{b_1}}} \omega \ge S_{QES}(B).
\end{equation}
So the strong subadditivity holds, which also indicates the non-negativity of integral measures.

\subsubsection{Monomorphism inequality}
\begin{figure}
	\centering
	\includegraphics[width=0.45\textwidth]{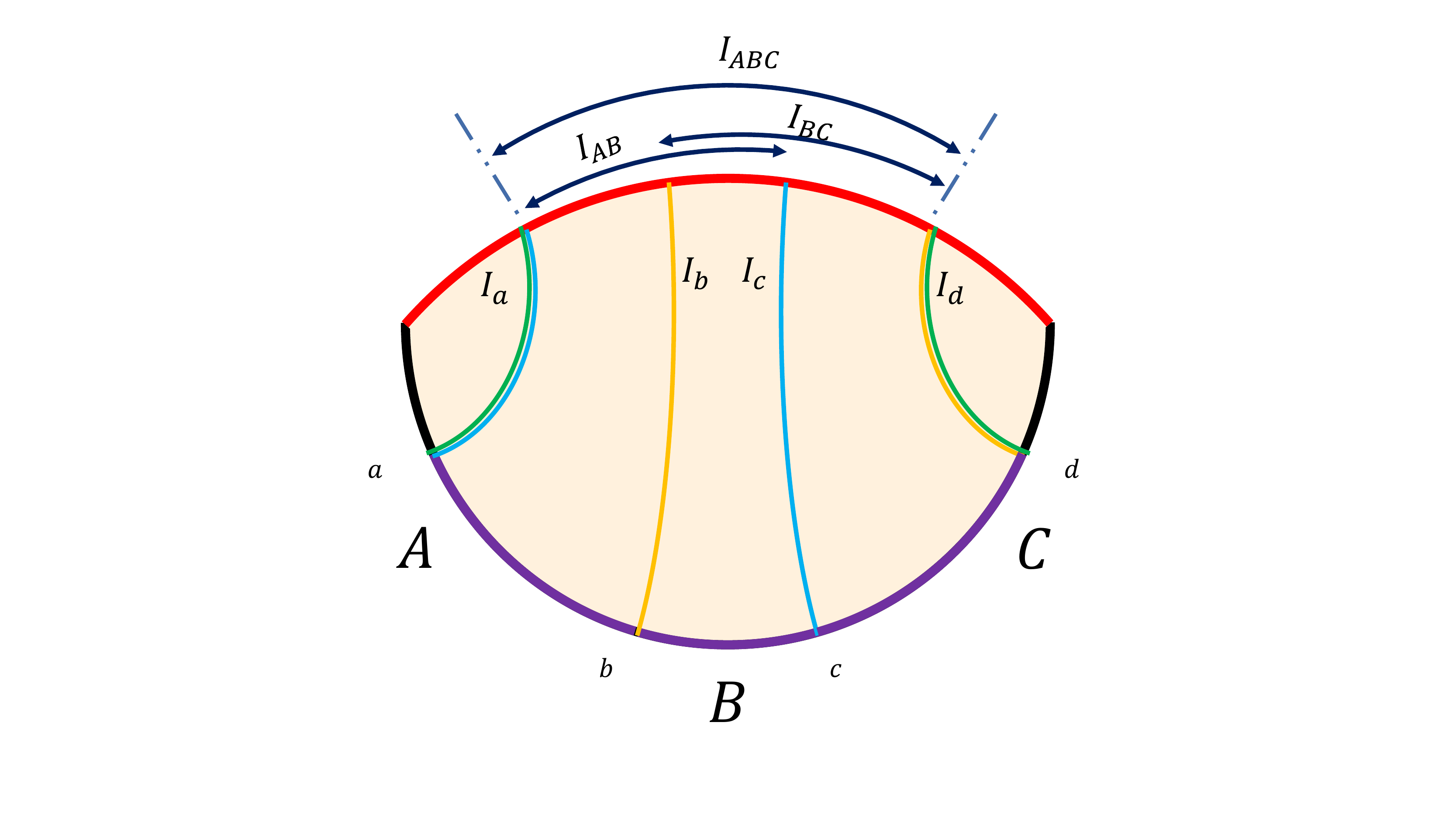}
	\includegraphics[width=0.45\textwidth]{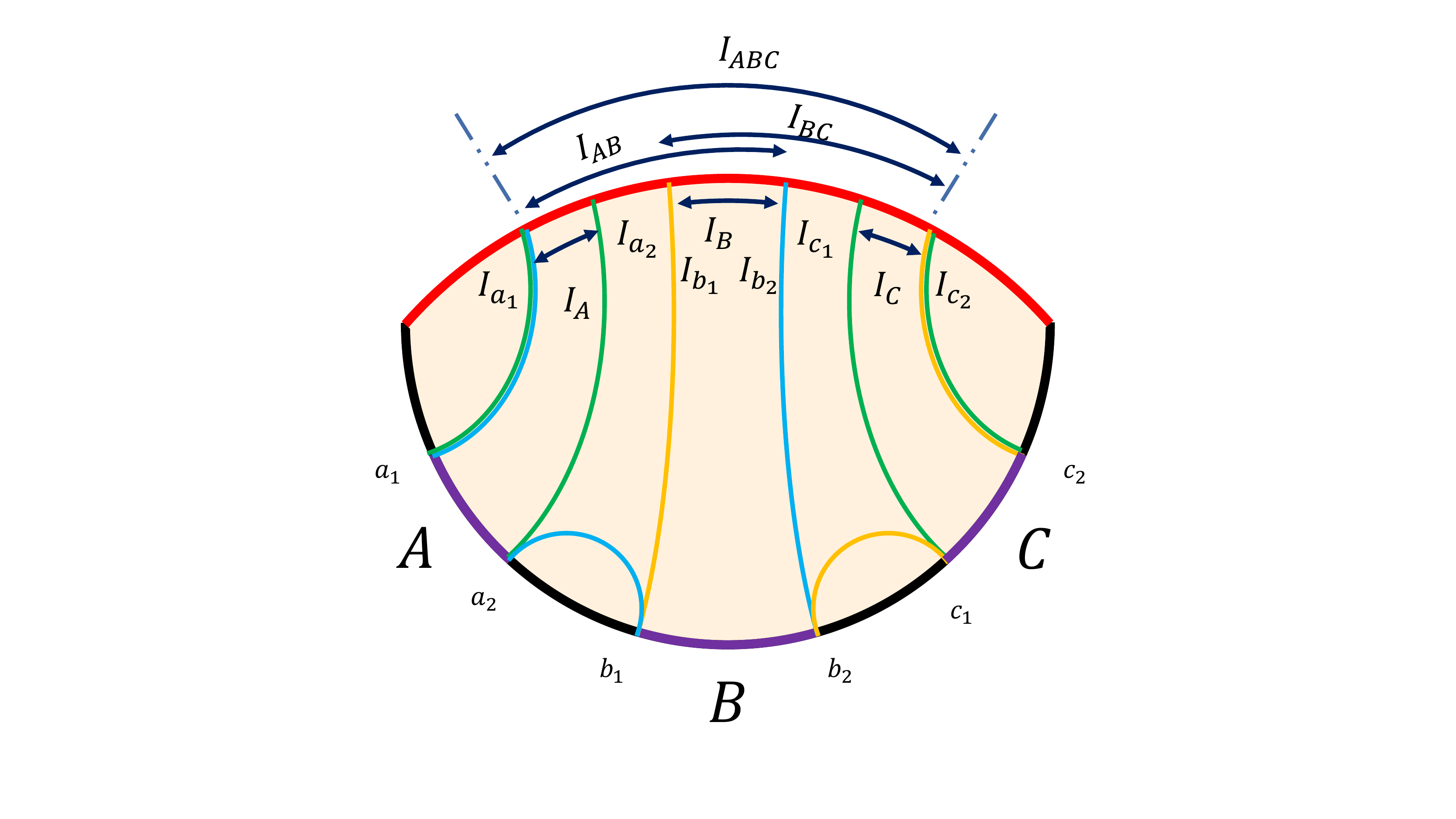}
	\caption{Left: The RT surfaces for interval $AB$ (the blue curves), $BC$ (the orange curves) and $AC$ (the green curves) as they are connected. Right: The RT surfaces for interval $AB$ (the blue curves), $BC$ (the orange curves) and $AC$ (the green curves) as they are disconnected.}
	\label{mmi}
\end{figure}

In this section, we will demonstrate the entanglement monotonicity of three subsystems $A$ , $B$ and $C$. We will consider two different types of subsystems: one where the three subsystems are connected together, and another where the three subsystems are not connected. It is easy to observe the monotonicity inequality of the entanglement entropy in the connected intervals, as shown on the left side of Fig. \ref{mmi}. Furthermore, we will show that the monotonicity holds for any three subsystems. The monotonicity inequality can be stated as follows:
\begin{equation}
	S_{\text{QES}}(AB)+S_{\text{QES}}(BC)+S_{\text{QES}}(AC)\ge S_{\text{QES}}(A)+S_{\text{QES}}(B)+S_{\text{QES}}(C)+S_{\text{QES}}(ABC).
\end{equation}

As shown in Fig. \ref{mmi}, the entanglement entropy of the disjoint interval $AC$ is determined by the sum of the entanglement entropy of interval $A$ and that of interval $C$. The entanglement entropy of interval $AB$ is determined by the lengths of the three geodesics shown in the figure, as well as the contributions of the islands corresponding to interval $AB$. Similarly, the entanglement entropy of interval $BC$ follows a similar pattern to that of interval $AB$.

\begin{equation}
	S_{\text{QES}}(AB)=\frac{A^{(d)}\left(\partial I_{AB}\right)}{4 G^{(d)}}+\frac{1}{4} \int_{p_{a_1} \Delta p_{I_{a_1}}} \omega+\frac{1}{4} \int_{p_{a_2} \Delta p_{b_1}} \omega+\frac{1}{4} \int_{p_{b_2} \Delta p_{I_{b_2}}} \omega,
\end{equation}
\begin{equation}
	S_{\text{QES}}(BC)=\frac{A^{(d)}\left(\partial I_{BC}\right)}{4 G^{(d)}}+\frac{1}{4} \int_{p_{b_1} \Delta p_{I_{b_1}}} \omega+\frac{1}{4} \int_{p_{b_2} \Delta p_{c_1}} \omega+\frac{1}{4} \int_{p_{c_2} \Delta p_{I_{c_2}}} \omega, 
\end{equation}
\begin{align}
	S_{\text{QES}}(AC)&=\frac{A^{(d)}\left(\partial I_{A}\right)}{4 G^{(d)}}+\frac{1}{4} \int_{p_{a_1} \Delta p_{I_{a_1}}} \omega+\frac{1}{4} \int_{p_{a_2} \Delta p_{I_{a_2}}} \omega\notag\\&+\frac{A^{(d)}\left(\partial I_{C}\right)}{4 G^{(d)}}+\frac{1}{4} \int_{p_{c_1} \Delta p_{I_{c_1}}} \omega+\frac{1}{4} \int_{p_{c_2} \Delta p_{I_{c_2}}} \omega.
\end{align}
The right sides of the above equations can be split and recombined to give entropies for different entangled regions, ensuring that the combined expression is not less than the entanglement entropy of the corresponding region, that is
\begin{equation}
	\frac{A^{(d)}\left(\partial I_{A}\right)}{4 G^{(d)}}+\frac{1}{4} \int_{p_{a_1} \Delta p_{I_{a_1}}} \omega+\frac{1}{4} \int_{p_{a_2} \Delta p_{I_{a_2}}} \omega=S_{\text{QES}}(A),
\end{equation}
\begin{equation}
	\frac{A^{(d)}\left(\partial I_{B}\right)}{4 G^{(d)}}+\frac{1}{4} \int_{p_{b_1} \Delta p_{I_{b_1}}} \omega+\frac{1}{4} \int_{p_{b_2} \Delta p_{I_{b_2}}} \omega\ge S_{Q E S}(B),
\end{equation}
\begin{equation}
	\frac{A^{(d)}\left(\partial I_{C}\right)}{4 G^{(d)}}+\frac{1}{4} \int_{p_{c_1} \Delta p_{I_{c_1}}} \omega+\frac{1}{4} \int_{p_{c_2} \Delta p_{I_{c_2}}} \omega=S_{Q E S}(C),
\end{equation}
\begin{equation}
	\frac{A^{(d)}\left(\partial I_{ABC}\right)}{4 G^{(d)}}+\frac{1}{4} \int_{p_{a_1} \Delta p_{I_{a_1}}} \omega+\frac{1}{4} \int_{p_{a_2} \Delta p_{b_1}} \omega+\frac{1}{4} \int_{p_{b_2} \Delta p_{I_{c_1}}} \omega+\frac{1}{4} \int_{p_{c_2} \Delta p_{I_{c_2}}} \omega\ge S_{Q E S}(ABC).
\end{equation}
This proves monogamy of holographic entanglement entropy.
\section{Relation between kinematics space and the contour function of entanglement entropy}
In \cite{Wen:2018whg}, the author proposed a conjecture regarding the contour function, stating that $s_{A}(A_{2})$ can be expressed as a linear combination of entanglement entropies of individual intervals inside $A$.
\begin{equation}
	s_{A}(A_{2} )=\frac{1}{2}(S_{A_{1}\cup A_{2}}+S_{A_{2}\cup A_{3}}-S_{A_{1}}-S_{A_{3}}), 
	\label{cos}
\end{equation}
Here, $A_{1}$, $A_{2}$, and $A_{3}$ correspond to the intervals $A, B$ and $C$, respectively, as discussed in the previous section 3. $s_{A}(A_{2})$ represents the contribution of interval $A_{2}$ to the entanglement entropy of the entire region. We will now demonstrate that the aforementioned conjecture is intuitive in the kinematic space and can be directly proven using it.
\begin{figure}
	\centering
	\includegraphics[width=0.45\textwidth]{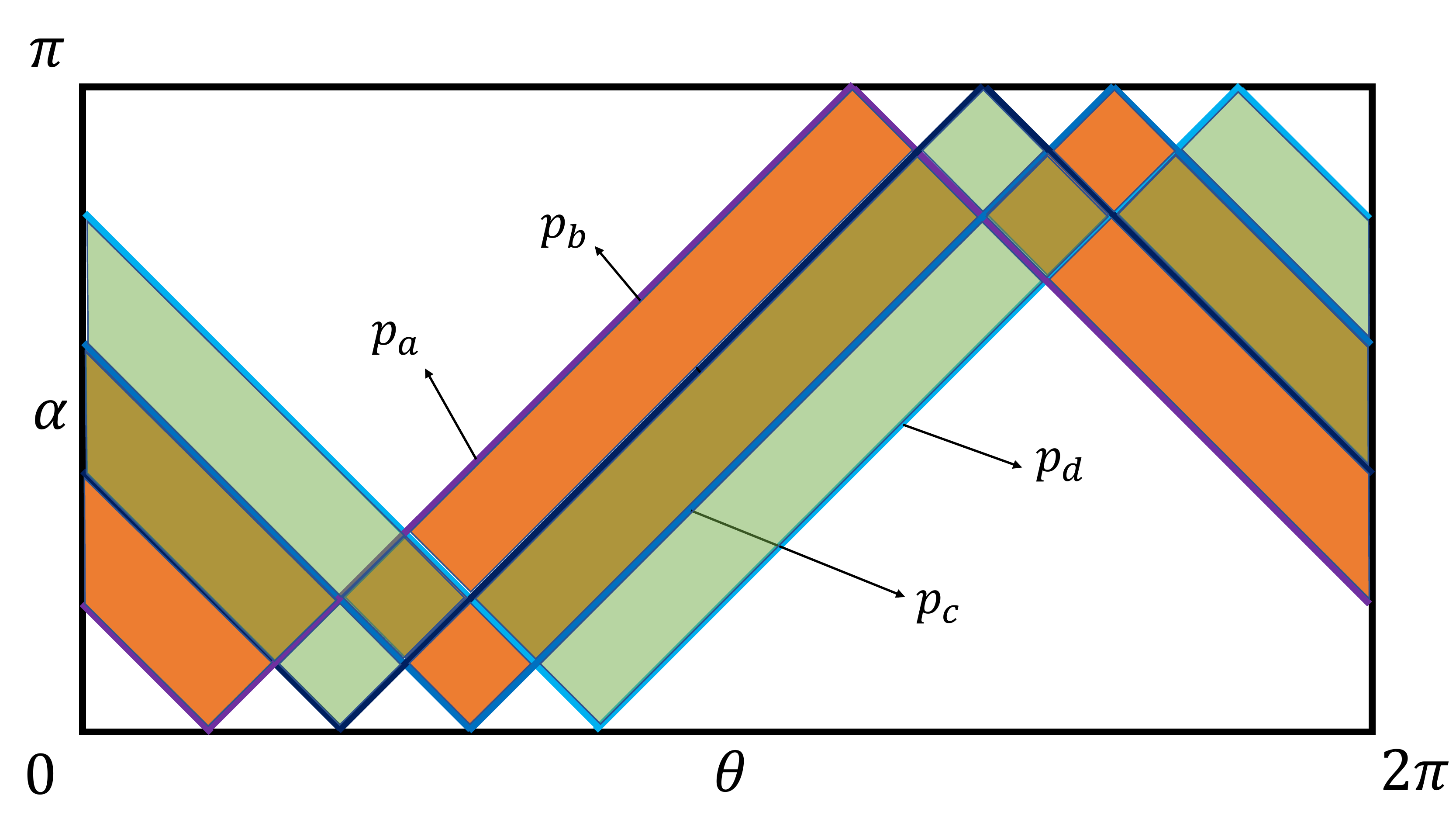}
	\includegraphics[width=0.45\textwidth]{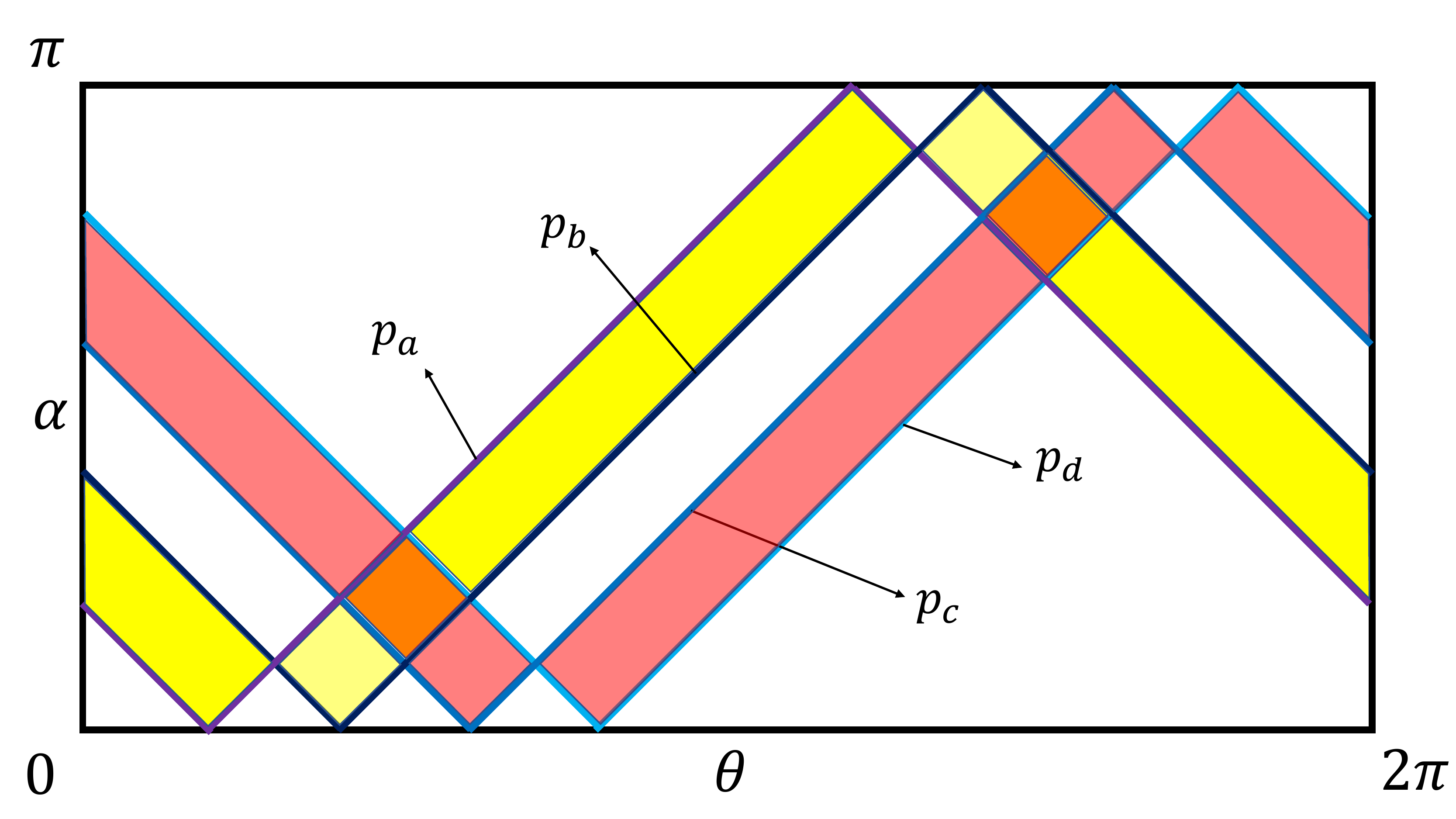}
	\caption{Left: integral regions for interval $A_{1}\cup A_{2}$ (the translucent green one) and $A_{2}\cup A_{3}$ (the orange one) in the kinematic space. Right: integral regions for interval $A_{1}$ (the pink one) and $A_{3}$ (the yellow one) in the kinematic space.}
	\label{fig12}
\end{figure}

As shown in Fig. \ref{fig12}, the entropy of the interval $A_{1}\cup A_{2}$ in the kinematic space is represented by the translucent green area, while the entropy of the interval $A_{2}\cup A_{3}$ is determined by integrating over the orange region
\begin{equation}
	S_{A_{1}\cup A_{2}}=\frac{1}{4} \int_{{p}_{a}\Delta  {p}_{c}}\omega,
\end{equation}
\begin{equation}
	S_{A_{2}\cup A_{3}}=\frac{1}{4} \int_{{p}_{b}\Delta  {p}_{d}}\omega.
\end{equation}
\begin{figure}
	\centering
	\includegraphics[width=0.45\textwidth]{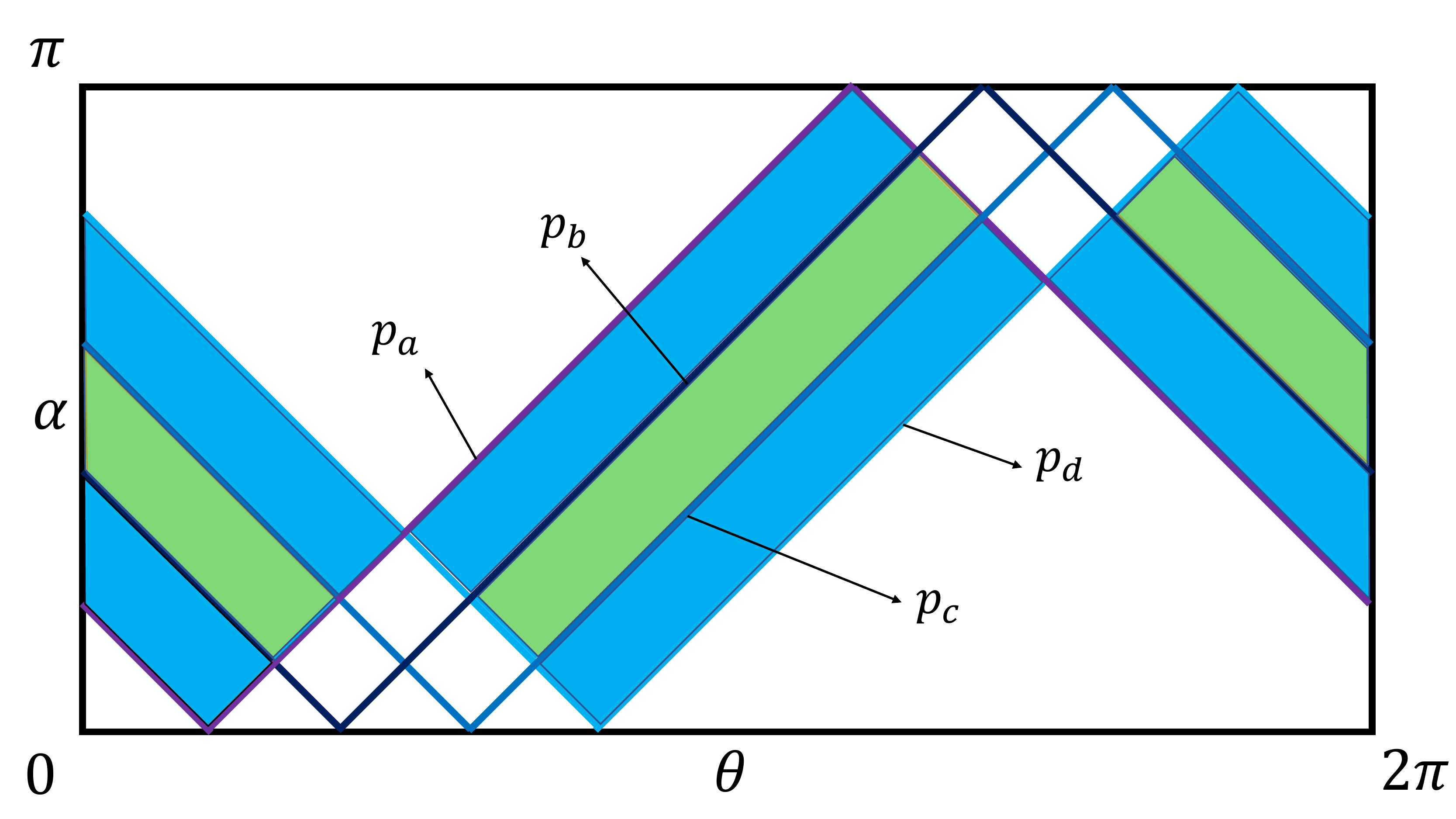}
	\caption{Contribution of interval $A_{2}$ to the entropy of the entire region.}
	\label{fig13}
\end{figure}
As shown in Fig. \ref{fig12}, the entanglement entropy of interval $A_{1}$ is determined by the pink region, while the entanglement entropy of interval $A_{3}$ is determined by integrating over the yellow region
\begin{equation}
	S_{A_{1}}=\frac{1}{4} \int_{{p}_{a}\Delta  {p}_{b}}\omega,
\end{equation}
\begin{equation}
	S_{A_{3}}=\frac{1}{4} \int_{{p}_{c}\Delta  {p}_{d}}\omega.
\end{equation}
As shown in Fig. \ref{fig13}, when combined with the entropy expression in the kinematic space mentioned above, the calculation result on the right hand side of Eq. \eqref{cos} is as follows:
\begin{equation}
\frac{1}{2}(S_{A_{1}\cup A_{2}}+S_{A_{2}\cup A_{3}}-S_{A_{1}}-S_{A_{3}})=\frac{1}{4} \int_{\text{green region}}\omega.
\end{equation}
The integral over the green region precisely quantifies the contribution of interval $A_{2}$ to the entanglement entropy of the entire region $A_{1}\cup A_{2}\cup A_{3}$.

\section{Conclusion}

In this paper, we first examine the holographic entanglement entropy inequality. Based on the relationship between integral geometry and holography, we establish the holographic entanglement entropy inequality for connected regions by comparing the integral regions in kinematic space. Additionally, we prove the holographic entanglement entropy inequality for any number of entangled regions from the perspective of double holography. In kinematic space, the entanglement entropy of a region can be expressed using geometric quantities within the bulk. These geometric quantities, such as length, can be derived from the integral quantities in kinematic space. This connection allows us to relate the concept of quantum extremum surfaces to kinematic space. Furthermore, we investigate the fine structure of entanglement entropy, which involves the so-called contour function. We show that employing kinematic space enables us to easily demonstrate a conjecture stating that the contribution of a region within the entangled region to the entanglement entropy of the entire region can be represented as a linear combination of the entanglement entropy of a single interval within the entangled region.

 %%%%%%%%%%%%%%%%%%%%%%%%%%%%%%%%%%%%
\section*{Acknowledgments}

This work is supported by the National Natural Science Foundation of China under Grant No. 11975116.

\end{document}